\documentclass[twocolumn,prl,twocolumn,superscriptaddress]{revtex4-1}
\usepackage{amsmath,amssymb,mathrsfs}
\usepackage{natbib}
\usepackage{subfigure}
\usepackage{tabularx}
\usepackage{epsfig}
\usepackage{longtable}
\usepackage{amsfonts}
\usepackage{rotating}
\usepackage{comment}
\usepackage{bbold}

\usepackage[unicode=true,bookmarks=true,bookmarksnumbered=false,bookmarksopen=false,breaklinks=false,pdfborder={0 0 1},backref=false,colorlinks=true]{hyperref}

\def\be{\begin{equation}}
\def\ee{\end{equation}}
\def\bea{\begin{eqnarray}}
\def\eea{\end{eqnarray}}

\def\up{\uparrow}
\def\down{\downarrow}

\def\blue{\color{blue}}

\hypersetup{linkcolor=magenta,urlcolor=blue,citecolor=blue,pdfstartview={FitH},hyperfootnotes=false,unicode=true}

\begin{document}
\title{The Sub-Exponential Critical Slowing Down at Floquet Time Crystal Phase Transition}

\author{Wenqian Zhang$^{\blue \dagger}$}
\affiliation{State Key Laboratory of Surface Physics, Key Laboratory of Micro and Nano Photonic Structures (MOE), and Department of Physics, Fudan University, Shanghai 200433, China}

\author{Yadong Wu$^{\blue \dagger}$} 
\affiliation{State Key Laboratory of Surface Physics, Key Laboratory of Micro and Nano Photonic Structures (MOE), and Department of Physics, Fudan University, Shanghai 200433, China}
\affiliation{Shanghai Qi Zhi Institute, AI Tower, Xuhui District, Shanghai 200232, China}

\author{Xingze Qiu}
\affiliation{State Key Laboratory of Surface Physics, Institute of Nanoelectronics and Quantum Computing, and Department of Physics, Fudan University, Shanghai 200438, China}   
\affiliation{Shanghai Qi Zhi Institute, AI Tower, Xuhui District, Shanghai 200232, China}

\author{Jue Nan} 
\email{juenan@fudan.edu.cn}
\affiliation{State Key Laboratory of Surface Physics, Institute of Nanoelectronics and Quantum Computing, and Department of Physics, Fudan University, Shanghai 200438, China} 
\affiliation{Shanghai Qi Zhi Institute, AI Tower, Xuhui District, Shanghai 200232, China}

\author{Xiaopeng Li}
\email{xiaopeng\underline{ }li@fudan.edu.cn}
\affiliation{State Key Laboratory of Surface Physics, Key Laboratory of Micro and Nano Photonic Structures (MOE), and Department of Physics, Fudan University, Shanghai 200433, China}
\affiliation{Institute for Nanoelectronic Devices and Quantum Computing, Fudan University, Shanghai 200433, China}
\affiliation{Shanghai Qi Zhi Institute, AI Tower, Xuhui District, Shanghai 200232, China}

\date{\today}

\begin{abstract}
{
Critical slowing down (CSD) has been a trademark of  critical dynamics for equilibrium phase transitions of a many-body system, where the relaxation time for the system to reach thermal equilibrium or quantum ground state diverges with system size. The time crystal phase transition has attracted much attention in recent years for it provides a scenario of phase transition of quantum dynamics, unlike conventional equilibrium phase transitions.  Here, we study critical dynamics near the Floquet time crystal phase transition. Its  critical behavior is described by introducing a space-time coarse grained correlation function, whose relaxation time diverges at the critical point revealing the CSD. 
This is demonstrated by investigating the Floquet dynamics of one-dimensional disordered spin chain. 
Through finite-size scaling analysis, we show the relaxation time has a universal sub-exponential scaling near the critical point, in sharp contrast to the standard power-law behavior for CSD in  equilibrium phase transitions. This prediction can be readily tested in present quantum simulation experiments.
} 
\end{abstract}
\maketitle

{\it Introduction.---}
Critical slowing down (CSD) is a ubiquitous phenomenon near phase transition reflecting a universal scaling relation between space and time, which emerges  in a broad range of thermodynamic systems such as electronic materials~\cite{2006_Lee_RMP,2015_Hartmann_PRL,2020_Kundu_PRL}, atomic quantum many-body systems~\cite{2021_Pan_NatPhys,2022_Pan_Science}, and even social science models ~\cite{2017_Billio_Financial,2022_Pirani_arXiv}. 
It has been introduced  in the Van Hove theory~\cite{1977_Hohenberg_ROMP,1954_Hove_PR,1954_Landau}, describing the existence of zero-relaxation-rate modes at a second order phase transition point. 
Its universality has been unveiled through a  phenomenological approach proposing a  dynamic scaling hypothesis~\cite{1967_Ferrell_PRL,1967_Halperin_PRL,1997_Soares_PRB}, as later justified by the renormalization group (RG) theory of critical dynamics~\cite{1972_Halperin_PRL,2013_Vosk_PRL}. 

In recent years, time crystal~\cite{2012_Wilczek_PRL,2015_Watanabe_PRL}, a dynamical phase has been attracting  tremendous research interests. 
It has been established in theory  that time crystal could be stabilized by long-range interactions~\cite{2019_Kozin_PRL}, disorder induced localization~\cite{2016_Khemani_PRL}, or intricate nonlinear effects~\cite{2018_Biao_PRL}. 
In experiments, it has been found in quantum many-body dynamics of a variety of quantum systems such as trapped ions~\cite{2017_Vishwanath_Nature}, superfluid Helium~\cite{2018_Autti_PRL,2021_Autti_Nature}, and cold atoms~\cite{2018_Smits_PRL}. 
Despite the extensive studies on the time crystal phases, the spontaneous symmetry breaking of the quantum dynamics near the phase transition remains less well understood.  In particular, for the time crystal phase transition is intrinsically a transition of dynamics, characterizing the CSD mechanism in this dynamical phase transition is fundamentally different from its equilibrium analogue~\cite{2021_Cai_PRL,2022_Cai_PRB,2022_Wei_arXiv}. 

In this work, we examine the spontaneous time crystal phase transition in the Floquet dynamics of a disordered spin chain in one dimension (1D)~\cite{2016_Khemani_PRL, 2016_Keyserlingk_PRB}. The CSD is extended from the equilibrium setting to the dynamical phase transition by performing coarse graining in both space and time dimensions.  We find the coarse grained dynamics exhibits a diverging relaxation time  near the dynamical phase transition. Based on the strong disorder RG theory~\cite{1992_Fisher_PRL,1995_Fisher_PRB,2013_Vosk_PRL,2014_Vosk_PRL,2018_Vasseur_PNAS}, we propose a finite-time-finite-size scaling form for the space-time coarse grained correlation. 
A large-scale numerical simulation for a system size up to $60$ spins is carried out by mapping the spin chain to Majorana fermions. 
The numerical results show a nice data-collapse on a universal curved surface. Our finite-size scaling analysis implies a sub-exponential slowing down for the critical dynamics, unlike  the standard CSD in equilibrium phase transitions. 

{\it Model.---}
A prominent scenario  to support the time crystal phase is the spontaneous period doubling of Floquet quantum dynamics of 1D disordered spin chains, where heating effects caused by periodic driving is suppressed by many-body localization~\cite{2015_Lazarides_PRL,2015_Ponte_PRL,2016_Dmitry_AoP}. 
The quantum dynamics is described by Floquet operators~\cite{2016_Khemani_PRL,2016_Keyserlingk_PRB,2018_Vasseur_PNAS}, 
\be
\hat{U}_F = 
\exp\left[{-i{t_2} \sum_{j=1}^{L-1} J_j \hat{\sigma}^z_j \hat{\sigma}^z_{j+1}}\right]
\exp\left[{-i{\frac{\pi}{2}t_1} g\sum_{j=1}^L \hat{\sigma}^x_j}\right] 
\label{eq:floquet_op}.
\ee
We take $t_1=t_2=1$ as a time unit, and choose a lognormal distribution for the random Ising couplings ($J_i$).  The couplings have a typical strength  $J_{\rm typ}$ given by  $\ln J_{\rm typ}=\frac{1}{L}\sum_j \ln J_j$, and their logarithms have a standard deviation $\sigma_J$. 
This model  exhibits spontaneous Ising symmetry breaking and discrete time translation symmetry breaking in the parameter regime  of  $1 -g <2 J_{\rm typ}/\pi<g\le 1$~\cite{2016_Khemani_PRL,2016_Keyserlingk_PRB}. Throughout this paper, we take $J_{\rm typ}=0.05 \pi$ and  $\sigma_J = 0.2 \pi$, for which the critical point separating the symmetry-broken time crystal and the symmetric phases is located  at $g_c=0.9$. Both of the phases are localized.

For each disorder configuration $\mathbf{J}\equiv(J_1, J_2, \ldots, J_L)$, we perform coarse-graining in both space and time directions by introducing 
\begin{align}
O(T,\mathbf{J}) &= \frac{1}{L L_t}\big |\sum_j \sum_{n=T+1}^{T+L_t} \langle \hat{\sigma}_j^z(n,\mathbf{J})\hat{\sigma}_j^z(0) \rangle (-1)^n \big |
\label{eq:order_parameter},
\end{align} 
where $\hat{\sigma}_j^z(n,\mathbf{J})=(\hat{U}_F(\mathbf{J})^\dag)^n \hat{\sigma}_j^z (\hat{U}_F(\mathbf{J}))^n$ and $L_t$ is the averaged number of Floquet periods. With the lattice sites and multiple consecutive Floquet periods averaged over,  the coarse-grained quantity  $O(T,{\bf J})$ is introduced to diagnose the collective relaxation dynamics of the system that develops CSD at the phase transition. 

In our numerical simulations, we choose a N\'eel state $|\up \down \up \down \cdots \rangle$ polarized in $z$ direction as the initial state of the Floquet evolution. 
The Floquet operator is mapped to dynamical evolution of non-interacting fermions by Jordan-Wigner transformation, by which the autocorrelation functions in Eq.~\eqref{eq:order_parameter}  are constructed from the Pfaffian of the fermion system ({\it Supplemental Materials}). 
In the symmetric paramagnetic phase, the spin polarization pattern in the initial state  would relax during the Floquet time evolution and eventually disappears at the long-time limit, which corresponds to a vanishing autocorrelation, $O(T, \mathbf{J})\to 0 $ at large $T$. 
In the time crystal phase, the initial spin polarization is retained in the quantum  dynamics even at the long-time limit. This is characterized by a finite autocorrelation, $O(T,\mathbf{J}) \neq 0$. 
The relaxation dynamics of $O(T,\mathbf{J})$ closely resembles the order parameter of Ising phase transition of equilibrium systems with a pinning field added to the boundary~\cite{2013_Assaad_PRX}. 

To extract the universal properties of the relaxation dynamics, we need to average over disorder configurations (different ${\bf J}$s). It has been established by RG analysis that different ways of averaging disorder would produce different critical scaling~\cite{1992_Fisher_PRL,1995_Fisher_PRB}.
Here we perform disorder averaging in two ways. 
One is arithmetic averaging and the other is geometric averaging, by which we obtain the mean value and the typical value, respectively, 
\begin{align}
    \textstyle \bar{A}^{\rm mea}& 
    \textstyle=\frac{1}{N_\mathbf{J}}\sum_{k=1}^{N_\mathbf{J}}A(\mathbf{J}_k),\label{mean}\\
    \textstyle \bar{A}^{\rm typ}& 
    \textstyle =\exp\left[\frac{1}{N_\mathbf{J}}\sum_{k=1}^{N_\mathbf{J}}\ln A(\mathbf{J}_k) \right],
    \label{typical}
\end{align}
where $A(\mathbf{J}_k)=\langle\hat{A}\rangle$ is the quantum state expectation of a physical observable $\hat{A}$, $k$ is the disorder index and $N_\mathbf{J}$ is the number of disorder configurations.

\begin{figure}
    \begin{center}
        \includegraphics[width=\linewidth]{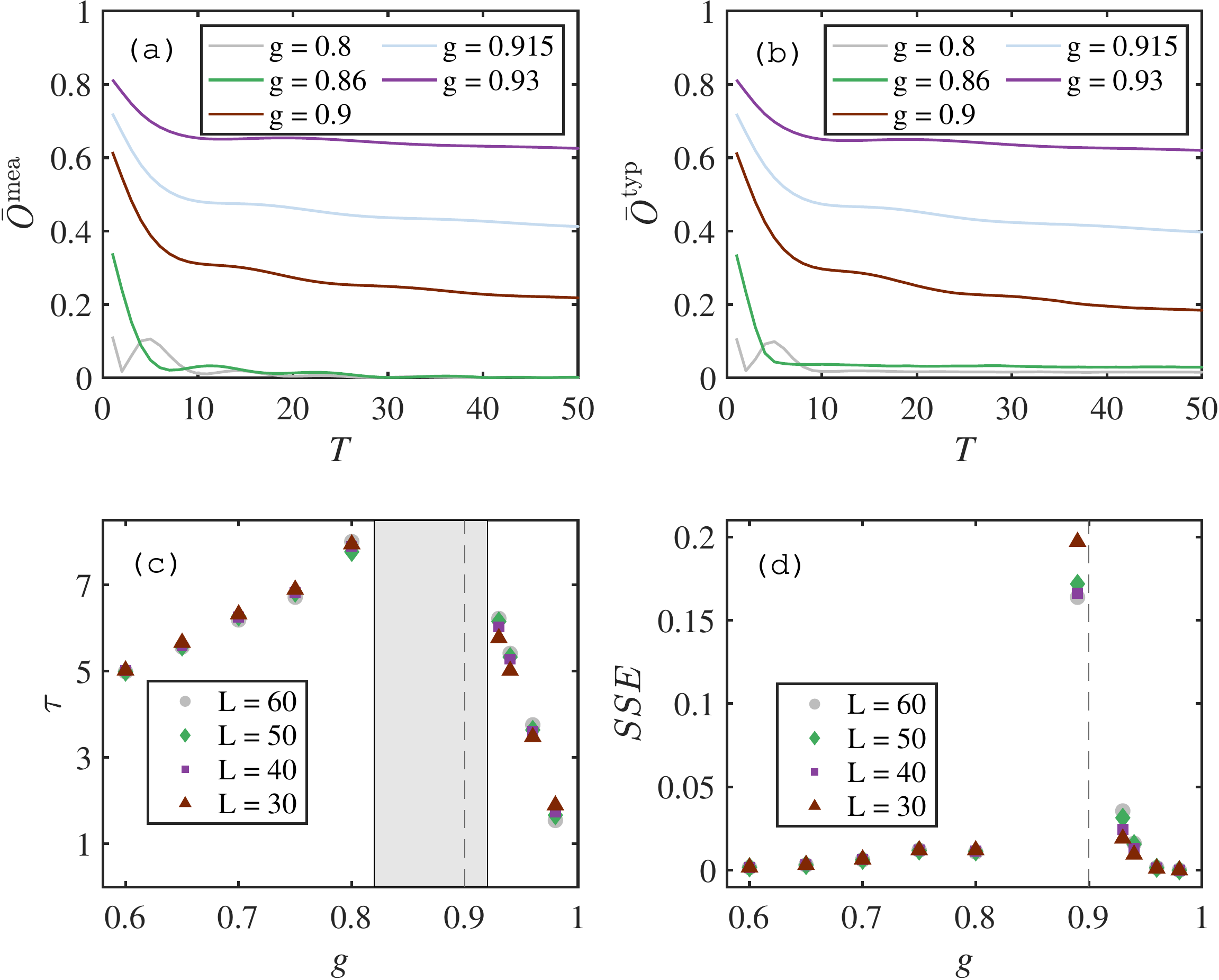}
    \end{center}
    \caption{The relaxation slowing down at critical point $g_c=0.9$. Early-time dynamics of (a) mean and (b) typical correlation (Eq.~\eqref{eq:order_parameter}). Here, we average over $L_t = 10$ Floquet periods and $3000$ disorder configurations. (c) The relaxation time obtained by fitting the relaxation dynamics to an exponential function. 
    It systematically increases as the tuning parameter $g$ approaches the critical point. The fitting fails in the shaded region, indicating non-trivial behavior of the critical dynamics.  
    (d) The sum of squares due to error ({\it SSE}) of the exponential fitting. 
    The dashed lines mark the critical point.} 
    \label{fig:1}
\end{figure}

{\it Relaxation across the Floquet time crystal phase transition.---} 
FIG.~\ref{fig:1} shows the relaxation dynamics. 
Away from the critical point, the system has  a  finite correlation length. It takes a finite amount of time for the system to dynamically relax. 
In the paramagnetic phase with $g<g_c$, we find $\overline{O(T)}^{\rm mea}$ and $\overline{O(T)}^{\rm typ}$ quickly decay to zero after brief oscillations. 
 The decay dynamics becomes relatively slower approaching the critical point. 
While in time crystal phase with $g>g_c$, $\overline{O(T)}^{\rm mea}$ and $\overline{O(T)}^{\rm typ}$ also undergo swift decay before reaching their static value. 
At the critical point $g=g_c$, the relaxation dynamics develops an apparent long-tail behavior, which implies the relaxation slows down dramatically.

\begin{figure*}
    \begin{center}
        \includegraphics[width=0.95\linewidth]{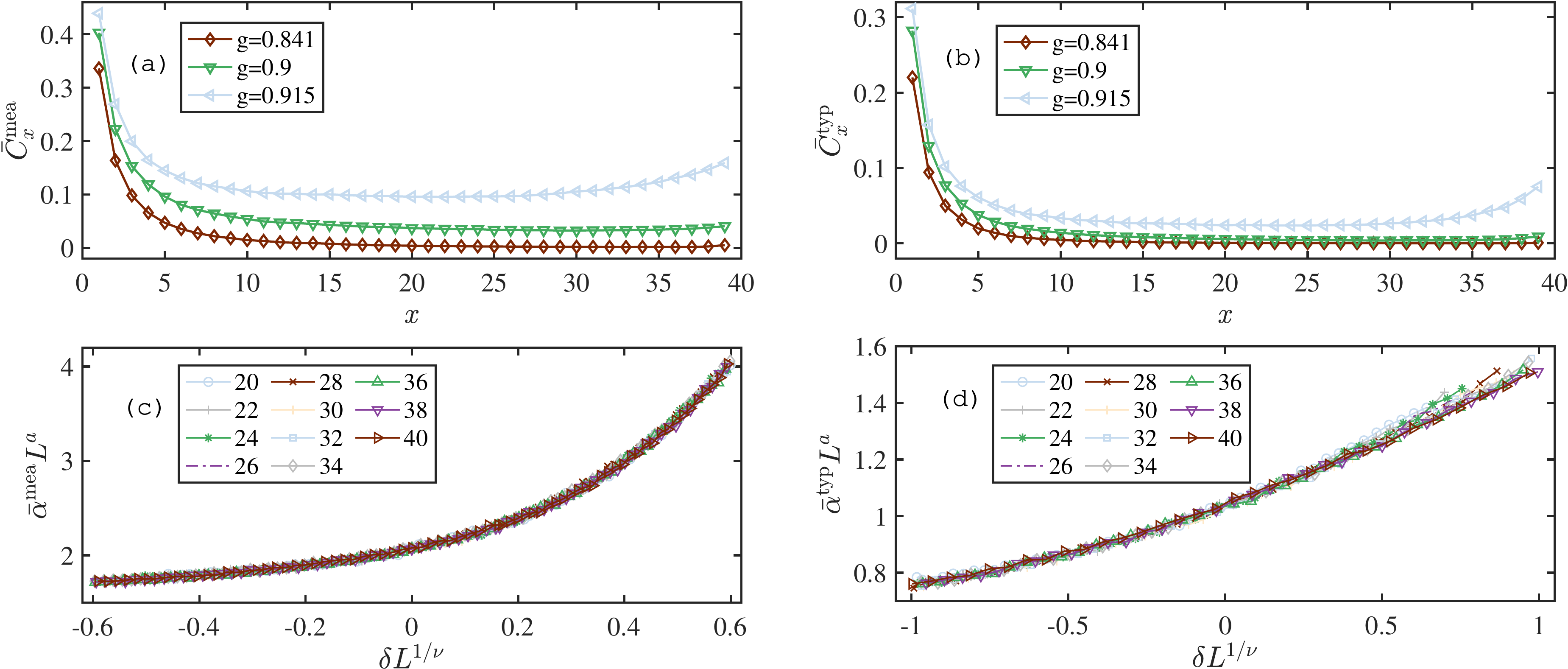}
    \end{center}
    \caption{Mean and typical  correlation functions of Floquet eigenstates.
    (a) Mean and (b) typical spin correlation functions with system size $L=40$ for the paramagnetic phase ($g = 0.841$), critical point ($g=0.9$), and time crystal phase ($g = 0.915$). Spin correlations are averaged over $2500$ disorder configurations.
    (c) and (d) show the mean and typical maximum eigenvalue $\bar{\alpha}$ of the correlation function matrices, respectively, which  acts as an  order parameter reflecting the spatial long range order. Finite-size scaling analysis has been performed on $\bar{\alpha}$. The correlation length exponent $\nu$, as defined by the correlation length divergence at critical point, $\xi \sim \delta^{-\nu}$, is $2.01\pm 0.02$ for the mean case (Eq.~\eqref{mean}) and $1.12\pm0.03$ for the typical case (Eq.~\eqref{typical}). The other exponent obtained for mean $\bar{\alpha}$ is ${\it{a}}= -0.170\pm 0.002$ whereas for typical $\bar{\alpha}$, ${\it{a}}= -0.512\pm 0.003$. 
    } 
    \label{fig:2}
\end{figure*}

From the relaxation dynamics, 
we extract the relaxation time $\tau$ by fitting the dynamics for different system sizes with an exponential function $\overline{O(T)}_{\rm fit}=a_1 e^{-T/\tau}+a_2$.  The dynamics of $ \overline{O(T)}^{\rm mea} $ fits well to the exponential function  in the regime not too close to the critical point. FIG.~\ref{fig:1} (c, d) show the  extracted  relaxation time and  the fitting errors, respectively. We observe that the relaxation time $\tau$ rises up near the critical point on both sides of the phase transition. Deep in the time crystal or the paramagnetic phase, the relaxation time is unaffected by the system size. In contrast, near the critical point, we find significant system-size dependence for the relaxation time. At the same time, the fitting error becomes  substantially larger. The results with geometric average are qualitatively similar ({\it see Supplemental Materials}). These observations indicate our introduced relaxation dynamics is indeed critical at the phase transition. 

{\it Finite-size scaling and dynamical criticality.---}
In order to systematically study the CSD of the relaxation near the phase transition, we analyze the Floquet dynamics with finite-size scaling theory. We first extract the $\nu$-exponents from the eigenstates of the effective Hamiltonian, which is defined by 
\begin{equation}
\hat{U}_F(\mathbf{J})=\exp[-i2\hat{H}_{\rm eff}(\mathbf{J})]. 
\end{equation}
FIG.~\ref{fig:2} (a, b) shows the disorder averaged correlations functions, $\overline{C_x}^{\rm mea}$ and $\overline{C_x}^{\rm typ}$ defined according to  Eq.~\eqref{mean} and Eq.~\eqref{typical} with  $C_x^k = |\langle \phi_k | \sigma^z_j \sigma^z_{j+x} |\phi_k \rangle|$.  Here we choose $j=\lceil (L-x)/2 \rceil, \ x=1,2,\cdots L-1$ to minimize boundary effects.  
In the paramagnetic phase, the correlation function decays exponentially down to zero with a finite correlation length, whereas in the time crystal phase, the correlation function saturates to a finite value at large distance. Near the critical point, the correlation length is comparable to the system size. The diverging behavior of the correlation length is reflected by the maximum eigenvalue ($\alpha_k$) of the correlation matrix $\mathbf{C}^k$ with matrix elements $\mathbf{C}^k_{ij} = \langle \phi_k | \hat{\sigma}^z_i \hat{\sigma}^z_j |\phi_k \rangle$~\cite{1965_Girardeau_Math}.  Its disorder averaged values, $\bar{\alpha}^{\rm mea}$ and $\bar{\alpha}^{\rm typ}$,   are introduced correspondingly. 
For systems with long-range order $\lim_{L\rightarrow \infty}\bar{\alpha}/L$ is finite, while $\lim_{L\rightarrow \infty}\bar{\alpha}/L \rightarrow 0$ for correlation functions that vanish at long distance. 
At the critical point, the $\bar{\alpha}$ value exhibits non-trivial scaling with the system size, reflecting the correlation-length criticality. 
With our numerical results, we find a reasonable data-collapse [Fig.~\ref{fig:2} (c, d)] by taking a finite-size scaling ansatz~\cite{1992_Fisher_PRL,1995_Fisher_PRB,2014_Vosk_PRL,2018_Vasseur_PNAS}, 
\begin{equation}
    \bar{\alpha}=L^{-a}f(\delta L^{1/\nu}),
    \label{nu}
\end{equation}
with $\delta=(\ln \frac{\pi}{2}g-\ln J_{\rm typ})/\sigma^2_J$. 
For arithmetic disorder average, we obtain $a = -0.170\pm 0.002$, and $\nu =2.01\pm0.02$.
For the typical average, we obtain $a=-0.512\pm0.003$ and $\nu = 1.12\pm0.03$, having a sizable difference from the arithmetic  average. 
These numerical results are consistent with disorder RG analysis at the infinite randomness fixed point~\cite{1995_Fisher_PRB}. 
 The typical correlation has a less divergent correlation length, which deviates substantially from the mean correlation for the latter receives significant contribution from distant resonant pairs~\cite{1992_Fisher_PRL}.

\begin{figure*}
    \begin{center}
        \includegraphics[width=\linewidth]{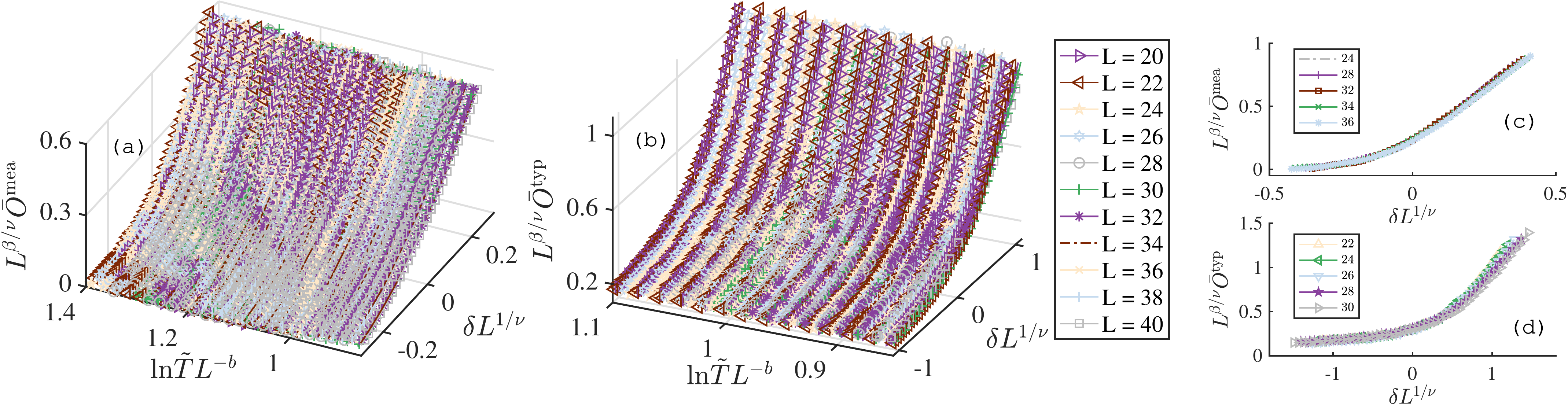}
    \end{center}
    \caption{The finite-size scaling results for the critical dynamics.  
    (a) shows the finite-size scaling for the mean value of the correlation in Eq.~\eqref{eq:order_parameter}. The fitting yields the exponents $b=0.45\pm 0.05,\ \beta=0.32\pm0.18,\ g_c = 0.90,\ T_0=2.03\pm 0.28$.    
    (b) shows the analysis for typical value of the correlation, where we find $b = 0.54\pm 0.09,\ \beta=0.37\pm 0.04, \ g_c = 0.90,$ and $T_0 = 1.49\pm 0.08$. 
    The fitting error is obtained by  bootstrapping. 
    We take the  $\nu$ exponents from the results  FIG.~\ref{fig:3}, to reduce the number of fitting parameters here. 
    Choosing the data for $\ln \tilde{T}L^{-b}\in [1,1.02]$,  cross-sections of the collapsed surface are shown in (c, d) for the mean and typical value of the correlation. 
    This demonstrates the high quality of the data-collapse in the finite-size scaling analysis. }
    \label{fig:3}
\end{figure*}

With the $\nu$-exponents obtained, we then perform the finite-size scaling analysis for the relaxation dynamics near the critical point. 
Although both of the spatial correlation length $\xi$ and the relaxation time $\tau$ have divergent behavior at the critical point, the disordered system lacks space-time symmetry---$\xi$ and $\tau$ exhibit different scalings.
Since the correlation length diverges at the critical point, it takes a divergent amount of time for the system to establish long-range spin correlations. 
It has been argued based on disorder RG theory that disordered systems obey an activated dynamic scaling $\ln \tau \sim \xi^b$ in a long time~\cite{1987_Fisher_JAP}. 
Since the dynamics of $\overline{O (T)}$ is analogous to the order parameter of an equilibrium system in presence of pinning fields, we propose a two-variable scaling function~\cite{1987_Fisher_JAP,2013_Assaad_PRX} 
\begin{equation}
    \overline{O(T)}=L^{-\beta/\nu} G (L/\xi, \ln \tilde{T}/\ln \tilde{\tau} ).
    \label{ansatz}
\end{equation}
Here we introduce the dimensionless time variable $\tilde{T}$ and $\tilde{\tau}$ defined as $T/T_0$ and $\tau/T_0$, where $T_0$ is some non-universal microscopic timescale analogous to  non-universal microscopic length scales in equilibrium phase transitions~\cite{2013_Vosk_PRL}.
Taking the activated dynamic scaling into account, the scaling function is rewritten as 
\begin{equation}
    \overline{O(T)}=L^{-\beta/\nu}f(\delta L^{1/\nu},\ln {\tilde{T}}/L^{b}). 
    \label{FTFS}
\end{equation}

For the coarse-graining, we average over multiple Floquet periods and choose $L_t = 10$ in Eq.~\eqref{eq:order_parameter}, to filter out the short-time dynamics. 
To reduce the contribution of non-universal dynamics from state initialization, it is required to examine the long-time limit, with ${\it T \gg L_t}$. In our numerical simulations, we choose {\it T}  in the range between $108$ and $806$ for finite-size scaling analysis.
We sample $1500$ disorder configurations and the results for the finite-size scaling with the two variables in Eq.~\eqref{FTFS} are shown in FIG.~\ref{fig:3} (a, b) for mean and typical value respectively. For both arithmetic and geometric averaging, the numerical data points from $11$ system sizes ranging between $L = 20$ and $L = 40$ collapse onto a single smooth curved surface with the critical exponents $b = 0.45 \pm0.05$ and $\beta =0.32\pm 0.18$ for mean value and $b = 0.54\pm 0.09$ and $\beta = 0.37\pm 0.04$ for typical value. 
The errors are obtained from bootstrapping~\cite{1979_Efron_SIAM}. 
The value of the exponent $b$ which defines the property of dynamic scaling shows the sub-exponential nature of critical relaxation. More concretely, for either way of disorder average, the CSD in this dynamical phase transition takes an approximate form of 
\begin{equation}
  {\tau} \propto \exp \left[ \sqrt{\delta ^{-\nu } } \right].
\end{equation} 
In the fitting we take the value of $\nu$-exponent determined through Eq.~\eqref{nu}  to reduce the number of the fitting variables here. 
The non-universal time scales are obtained to be  $T_0 = 2.0\pm0.3$ for mean value and $T_0 = 1.49\pm 0.08$ for typical value. Considering the evolution time we consider, this microscopic time scale is relatively small.

We further illustrate the quality of two-dimensional data-collapse by cutting out a slice of data points, which are shown by line plots in FIG.~\ref{fig:3} (c, d). Here, we pick data points in a thin slice with $\ln \tilde{T} L^{-b} \in [1.00, 1.02]$ for both mean and typical correlation and plot $L^{\beta/\nu}\bar{O}$ as a function  of  $\delta L^{1/\nu}$. As shown in FIG.~\ref{fig:3} (c, d), the data points obtained from various system sizes consistently fall on a smooth curve taking the determined exponents, which yields a nice one-dimensional data-collapse as widely used in analyzing equilibrium phase transitions~\cite{1999_Newman_Oxford}.  
This confirms the high quality of the two-dimensional data-collapse, and further justifies the scaling form assumed in Eq.~\eqref{ansatz}. The analysis also verifies the critical exponents we determine to describe the CSD, and the resultant sub-exponential relation of the relaxation time with system size.

{\it Conclusion.---}
{
We find  the phenomenon of critical slowing down widely studied in equilibrium phase transitions also carries over to the dynamical Floquet time crystal phase transition. 
This is demonstrated by large-scale simulation of a disordered quantum spin chain. The critical slowing down is described by introducing a space-time coarse-grained spin correlation, which can be measured directly in quantum simulation experiments. Through finite-size scaling analysis, we show the relaxation time of the coarse-grained spin correlation  has a universal divergence near the critical point. With both arithmetic and geometric average, the relaxation time has an approximate sub-exponential form, which implies drastic critical slowing down of the time-crystal phase transition. 
}

{\it Acknowledgement.---} 
We acknowledge Xiong-Jun Liu, Biao Huang, Zi Cai and W. Vincent Liu for helpful discussion. 
This work is supported by National Program on Key Basic Research Project of China (Grant No. 2021YFA1400900), National Natural Science Foundation of China (Grants No. 11934002), Shanghai Municipal Science and Technology Major Project (Grant No. 2019SHZDZX01), and  Shanghai Science Foundation (Grants No.21QA1400500). 

$^{\blue \dagger}$ These authors contributed equally to this work. 

\bibliography{reference}

\begin{thebibliography}{40}%
\makeatletter
\providecommand \@ifxundefined [1]{%
 \@ifx{#1\undefined}
}%
\providecommand \@ifnum [1]{%
 \ifnum #1\expandafter \@firstoftwo
 \else \expandafter \@secondoftwo
 \fi
}%
\providecommand \@ifx [1]{%
 \ifx #1\expandafter \@firstoftwo
 \else \expandafter \@secondoftwo
 \fi
}%
\providecommand \natexlab [1]{#1}%
\providecommand \enquote  [1]{``#1''}%
\providecommand \bibnamefont  [1]{#1}%
\providecommand \bibfnamefont [1]{#1}%
\providecommand \citenamefont [1]{#1}%
\providecommand \href@noop [0]{\@secondoftwo}%
\providecommand \href [0]{\begingroup \@sanitize@url \@href}%
\providecommand \@href[1]{\@@startlink{#1}\@@href}%
\providecommand \@@href[1]{\endgroup#1\@@endlink}%
\providecommand \@sanitize@url [0]{\catcode `\\12\catcode `\$12\catcode
  `\&12\catcode `\#12\catcode `\^12\catcode `\_12\catcode `\%12\relax}%
\providecommand \@@startlink[1]{}%
\providecommand \@@endlink[0]{}%
\providecommand \url  [0]{\begingroup\@sanitize@url \@url }%
\providecommand \@url [1]{\endgroup\@href {#1}{\urlprefix }}%
\providecommand \urlprefix  [0]{URL }%
\providecommand \Eprint [0]{\href }%
\providecommand \doibase [0]{http://dx.doi.org/}%
\providecommand \selectlanguage [0]{\@gobble}%
\providecommand \bibinfo  [0]{\@secondoftwo}%
\providecommand \bibfield  [0]{\@secondoftwo}%
\providecommand \translation [1]{[#1]}%
\providecommand \BibitemOpen [0]{}%
\providecommand \bibitemStop [0]{}%
\providecommand \bibitemNoStop [0]{.\EOS\space}%
\providecommand \EOS [0]{\spacefactor3000\relax}%
\providecommand \BibitemShut  [1]{\csname bibitem#1\endcsname}%
\let\auto@bib@innerbib\@empty
\bibitem [{\citenamefont {Lee}\ \emph {et~al.}(2006)\citenamefont {Lee},
  \citenamefont {Nagaosa},\ and\ \citenamefont {Wen}}]{2006_Lee_RMP}%
  \BibitemOpen
  \bibfield  {author} {\bibinfo {author} {\bibfnamefont {P.~A.}\ \bibnamefont
  {Lee}}, \bibinfo {author} {\bibfnamefont {N.}~\bibnamefont {Nagaosa}}, \ and\
  \bibinfo {author} {\bibfnamefont {X.-G.}\ \bibnamefont {Wen}},\ }\href
  {\doibase 10.1103/RevModPhys.78.17} {\bibfield  {journal} {\bibinfo
  {journal} {Rev. Mod. Phys.}\ }\textbf {\bibinfo {volume} {78}},\ \bibinfo
  {pages} {17} (\bibinfo {year} {2006})}\BibitemShut {NoStop}%
\bibitem [{\citenamefont {Hartmann}\ \emph {et~al.}(2015)\citenamefont
  {Hartmann}, \citenamefont {Zielke}, \citenamefont {Polzin}, \citenamefont
  {Sasaki},\ and\ \citenamefont {M\"uller}}]{2015_Hartmann_PRL}%
  \BibitemOpen
  \bibfield  {author} {\bibinfo {author} {\bibfnamefont {B.}~\bibnamefont
  {Hartmann}}, \bibinfo {author} {\bibfnamefont {D.}~\bibnamefont {Zielke}},
  \bibinfo {author} {\bibfnamefont {J.}~\bibnamefont {Polzin}}, \bibinfo
  {author} {\bibfnamefont {T.}~\bibnamefont {Sasaki}}, \ and\ \bibinfo {author}
  {\bibfnamefont {J.}~\bibnamefont {M\"uller}},\ }\href {\doibase
  10.1103/PhysRevLett.114.216403} {\bibfield  {journal} {\bibinfo  {journal}
  {Phys. Rev. Lett.}\ }\textbf {\bibinfo {volume} {114}},\ \bibinfo {pages}
  {216403} (\bibinfo {year} {2015})}\BibitemShut {NoStop}%
\bibitem [{\citenamefont {Kundu}\ \emph {et~al.}(2020)\citenamefont {Kundu},
  \citenamefont {Bar}, \citenamefont {Nayak},\ and\ \citenamefont
  {Bansal}}]{2020_Kundu_PRL}%
  \BibitemOpen
  \bibfield  {author} {\bibinfo {author} {\bibfnamefont {S.}~\bibnamefont
  {Kundu}}, \bibinfo {author} {\bibfnamefont {T.}~\bibnamefont {Bar}}, \bibinfo
  {author} {\bibfnamefont {R.~K.}\ \bibnamefont {Nayak}}, \ and\ \bibinfo
  {author} {\bibfnamefont {B.}~\bibnamefont {Bansal}},\ }\href {\doibase
  10.1103/PhysRevLett.124.095703} {\bibfield  {journal} {\bibinfo  {journal}
  {Phys. Rev. Lett.}\ }\textbf {\bibinfo {volume} {124}},\ \bibinfo {pages}
  {095703} (\bibinfo {year} {2020})}\BibitemShut {NoStop}%
\bibitem [{\citenamefont {Sun}\ \emph {et~al.}(2021)\citenamefont {Sun},
  \citenamefont {Yang}, \citenamefont {Wang}, \citenamefont {Zhou},
  \citenamefont {Su}, \citenamefont {Dai}, \citenamefont {Yuan},\ and\
  \citenamefont {Pan}}]{2021_Pan_NatPhys}%
  \BibitemOpen
  \bibfield  {author} {\bibinfo {author} {\bibfnamefont {H.}~\bibnamefont
  {Sun}}, \bibinfo {author} {\bibfnamefont {B.}~\bibnamefont {Yang}}, \bibinfo
  {author} {\bibfnamefont {H.-Y.}\ \bibnamefont {Wang}}, \bibinfo {author}
  {\bibfnamefont {Z.-Y.}\ \bibnamefont {Zhou}}, \bibinfo {author}
  {\bibfnamefont {G.-X.}\ \bibnamefont {Su}}, \bibinfo {author} {\bibfnamefont
  {H.-N.}\ \bibnamefont {Dai}}, \bibinfo {author} {\bibfnamefont {Z.-S.}\
  \bibnamefont {Yuan}}, \ and\ \bibinfo {author} {\bibfnamefont {J.-W.}\
  \bibnamefont {Pan}},\ }\href@noop {} {\bibfield  {journal} {\bibinfo
  {journal} {Nature Physics}\ }\textbf {\bibinfo {volume} {17}},\ \bibinfo
  {pages} {990} (\bibinfo {year} {2021})}\BibitemShut {NoStop}%
\bibitem [{\citenamefont {Li}\ \emph {et~al.}(2022)\citenamefont {Li},
  \citenamefont {Luo}, \citenamefont {Wang}, \citenamefont {Xie}, \citenamefont
  {Liu}, \citenamefont {Hu}, \citenamefont {Chen}, \citenamefont {Yao},\ and\
  \citenamefont {Pan}}]{2022_Pan_Science}%
  \BibitemOpen
  \bibfield  {author} {\bibinfo {author} {\bibfnamefont {X.}~\bibnamefont
  {Li}}, \bibinfo {author} {\bibfnamefont {X.}~\bibnamefont {Luo}}, \bibinfo
  {author} {\bibfnamefont {S.}~\bibnamefont {Wang}}, \bibinfo {author}
  {\bibfnamefont {K.}~\bibnamefont {Xie}}, \bibinfo {author} {\bibfnamefont
  {X.-P.}\ \bibnamefont {Liu}}, \bibinfo {author} {\bibfnamefont
  {H.}~\bibnamefont {Hu}}, \bibinfo {author} {\bibfnamefont {Y.-A.}\
  \bibnamefont {Chen}}, \bibinfo {author} {\bibfnamefont {X.-C.}\ \bibnamefont
  {Yao}}, \ and\ \bibinfo {author} {\bibfnamefont {J.-W.}\ \bibnamefont
  {Pan}},\ }\href@noop {} {\bibfield  {journal} {\bibinfo  {journal} {Science}\
  }\textbf {\bibinfo {volume} {375}},\ \bibinfo {pages} {528} (\bibinfo {year}
  {2022})}\BibitemShut {NoStop}%
\bibitem [{\citenamefont {Gatfaoui}\ \emph {et~al.}(2017)\citenamefont
  {Gatfaoui}, \citenamefont {Nagot},\ and\ \citenamefont {{de
  Peretti}}}]{2017_Billio_Financial}%
  \BibitemOpen
  \bibfield  {author} {\bibinfo {author} {\bibfnamefont {H.}~\bibnamefont
  {Gatfaoui}}, \bibinfo {author} {\bibfnamefont {I.}~\bibnamefont {Nagot}}, \
  and\ \bibinfo {author} {\bibfnamefont {P.}~\bibnamefont {{de Peretti}}},\
  }in\ \href {\doibase https://doi.org/10.1016/B978-1-78548-085-0.50003-0}
  {\emph {\bibinfo {booktitle} {Systemic Risk Tomography}}},\ \bibinfo {editor}
  {edited by\ \bibinfo {editor} {\bibfnamefont {M.}~\bibnamefont {Billio}},
  \bibinfo {editor} {\bibfnamefont {L.}~\bibnamefont {Pelizzon}}, \ and\
  \bibinfo {editor} {\bibfnamefont {R.}~\bibnamefont {Savona}}}\ (\bibinfo
  {publisher} {Elsevier},\ \bibinfo {year} {2017})\ pp.\ \bibinfo {pages}
  {73--93}\BibitemShut {NoStop}%
\bibitem [{\citenamefont {Pirani}\ and\ \citenamefont
  {Jafarpour}(2022)}]{2022_Pirani_arXiv}%
  \BibitemOpen
  \bibfield  {author} {\bibinfo {author} {\bibfnamefont {M.}~\bibnamefont
  {Pirani}}\ and\ \bibinfo {author} {\bibfnamefont {S.}~\bibnamefont
  {Jafarpour}},\ }\href {\doibase 10.48550/ARXIV.2208.03881} {\enquote
  {\bibinfo {title} {Network critical slowing down: Data-driven detection of
  critical transitions in nonlinear networks},}\ } (\bibinfo {year}
  {2022})\BibitemShut {NoStop}%
\bibitem [{\citenamefont {Hohenberg}\ and\ \citenamefont
  {Halperin}(1977)}]{1977_Hohenberg_ROMP}%
  \BibitemOpen
  \bibfield  {author} {\bibinfo {author} {\bibfnamefont {P.~C.}\ \bibnamefont
  {Hohenberg}}\ and\ \bibinfo {author} {\bibfnamefont {B.~I.}\ \bibnamefont
  {Halperin}},\ }\href {\doibase 10.1103/RevModPhys.49.435} {\bibfield
  {journal} {\bibinfo  {journal} {Rev. Mod. Phys.}\ }\textbf {\bibinfo {volume}
  {49}},\ \bibinfo {pages} {435} (\bibinfo {year} {1977})}\BibitemShut
  {NoStop}%
\bibitem [{\citenamefont {Van~Hove}(1954)}]{1954_Hove_PR}%
  \BibitemOpen
  \bibfield  {author} {\bibinfo {author} {\bibfnamefont {L.}~\bibnamefont
  {Van~Hove}},\ }\href {\doibase 10.1103/PhysRev.95.1374} {\bibfield  {journal}
  {\bibinfo  {journal} {Phys. Rev.}\ }\textbf {\bibinfo {volume} {95}},\
  \bibinfo {pages} {1374} (\bibinfo {year} {1954})}\BibitemShut {NoStop}%
\bibitem [{\citenamefont {Landau}\ and\ \citenamefont
  {Khalatnikov}(1954)}]{1954_Landau}%
  \BibitemOpen
  \bibfield  {author} {\bibinfo {author} {\bibfnamefont {L.}~\bibnamefont
  {Landau}}\ and\ \bibinfo {author} {\bibfnamefont {I.}~\bibnamefont
  {Khalatnikov}},\ }\href@noop {} {\bibfield  {journal} {\bibinfo  {journal}
  {Dokladu Akademii Nauk, SSSR}\ }\textbf {\bibinfo {volume} {96}},\ \bibinfo
  {pages} {469} (\bibinfo {year} {1954})}\BibitemShut {NoStop}%
\bibitem [{\citenamefont {Ferrell}\ \emph {et~al.}(1967)\citenamefont
  {Ferrell}, \citenamefont {Menyh\'ard}, \citenamefont {Schmidt}, \citenamefont
  {Schwabl},\ and\ \citenamefont {Sz\'epfalusy}}]{1967_Ferrell_PRL}%
  \BibitemOpen
  \bibfield  {author} {\bibinfo {author} {\bibfnamefont {R.~A.}\ \bibnamefont
  {Ferrell}}, \bibinfo {author} {\bibfnamefont {N.}~\bibnamefont {Menyh\'ard}},
  \bibinfo {author} {\bibfnamefont {H.}~\bibnamefont {Schmidt}}, \bibinfo
  {author} {\bibfnamefont {F.}~\bibnamefont {Schwabl}}, \ and\ \bibinfo
  {author} {\bibfnamefont {P.}~\bibnamefont {Sz\'epfalusy}},\ }\href {\doibase
  10.1103/PhysRevLett.18.891} {\bibfield  {journal} {\bibinfo  {journal} {Phys.
  Rev. Lett.}\ }\textbf {\bibinfo {volume} {18}},\ \bibinfo {pages} {891}
  (\bibinfo {year} {1967})}\BibitemShut {NoStop}%
\bibitem [{\citenamefont {Halperin}\ and\ \citenamefont
  {Hohenberg}(1967)}]{1967_Halperin_PRL}%
  \BibitemOpen
  \bibfield  {author} {\bibinfo {author} {\bibfnamefont {B.~I.}\ \bibnamefont
  {Halperin}}\ and\ \bibinfo {author} {\bibfnamefont {P.~C.}\ \bibnamefont
  {Hohenberg}},\ }\href {\doibase 10.1103/PhysRevLett.19.700} {\bibfield
  {journal} {\bibinfo  {journal} {Phys. Rev. Lett.}\ }\textbf {\bibinfo
  {volume} {19}},\ \bibinfo {pages} {700} (\bibinfo {year} {1967})}\BibitemShut
  {NoStop}%
\bibitem [{\citenamefont {Silv\'erio~Soares}\ \emph {et~al.}(1997)\citenamefont
  {Silv\'erio~Soares}, \citenamefont {Kamphorst Leal~da Silva},\ and\
  \citenamefont {S\'aBarreto}}]{1997_Soares_PRB}%
  \BibitemOpen
  \bibfield  {author} {\bibinfo {author} {\bibfnamefont {M.}~\bibnamefont
  {Silv\'erio~Soares}}, \bibinfo {author} {\bibfnamefont {J.}~\bibnamefont
  {Kamphorst Leal~da Silva}}, \ and\ \bibinfo {author} {\bibfnamefont {F.~C.}\
  \bibnamefont {S\'aBarreto}},\ }\href {\doibase 10.1103/PhysRevB.55.1021}
  {\bibfield  {journal} {\bibinfo  {journal} {Phys. Rev. B}\ }\textbf {\bibinfo
  {volume} {55}},\ \bibinfo {pages} {1021} (\bibinfo {year}
  {1997})}\BibitemShut {NoStop}%
\bibitem [{\citenamefont {Halperin}\ \emph {et~al.}(1972)\citenamefont
  {Halperin}, \citenamefont {Hohenberg},\ and\ \citenamefont
  {Ma}}]{1972_Halperin_PRL}%
  \BibitemOpen
  \bibfield  {author} {\bibinfo {author} {\bibfnamefont {B.~I.}\ \bibnamefont
  {Halperin}}, \bibinfo {author} {\bibfnamefont {P.~C.}\ \bibnamefont
  {Hohenberg}}, \ and\ \bibinfo {author} {\bibfnamefont {S.-k.}\ \bibnamefont
  {Ma}},\ }\href {\doibase 10.1103/PhysRevLett.29.1548} {\bibfield  {journal}
  {\bibinfo  {journal} {Phys. Rev. Lett.}\ }\textbf {\bibinfo {volume} {29}},\
  \bibinfo {pages} {1548} (\bibinfo {year} {1972})}\BibitemShut {NoStop}%
\bibitem [{\citenamefont {Vosk}\ and\ \citenamefont
  {Altman}(2013)}]{2013_Vosk_PRL}%
  \BibitemOpen
  \bibfield  {author} {\bibinfo {author} {\bibfnamefont {R.}~\bibnamefont
  {Vosk}}\ and\ \bibinfo {author} {\bibfnamefont {E.}~\bibnamefont {Altman}},\
  }\href {\doibase 10.1103/PhysRevLett.110.067204} {\bibfield  {journal}
  {\bibinfo  {journal} {Phys. Rev. Lett.}\ }\textbf {\bibinfo {volume} {110}},\
  \bibinfo {pages} {067204} (\bibinfo {year} {2013})}\BibitemShut {NoStop}%
\bibitem [{\citenamefont {Wilczek}(2012)}]{2012_Wilczek_PRL}%
  \BibitemOpen
  \bibfield  {author} {\bibinfo {author} {\bibfnamefont {F.}~\bibnamefont
  {Wilczek}},\ }\href {\doibase 10.1103/PhysRevLett.109.160401} {\bibfield
  {journal} {\bibinfo  {journal} {Phys. Rev. Lett.}\ }\textbf {\bibinfo
  {volume} {109}},\ \bibinfo {pages} {160401} (\bibinfo {year}
  {2012})}\BibitemShut {NoStop}%
\bibitem [{\citenamefont {Watanabe}\ and\ \citenamefont
  {Oshikawa}(2015)}]{2015_Watanabe_PRL}%
  \BibitemOpen
  \bibfield  {author} {\bibinfo {author} {\bibfnamefont {H.}~\bibnamefont
  {Watanabe}}\ and\ \bibinfo {author} {\bibfnamefont {M.}~\bibnamefont
  {Oshikawa}},\ }\href {\doibase 10.1103/PhysRevLett.114.251603} {\bibfield
  {journal} {\bibinfo  {journal} {Phys. Rev. Lett.}\ }\textbf {\bibinfo
  {volume} {114}},\ \bibinfo {pages} {251603} (\bibinfo {year}
  {2015})}\BibitemShut {NoStop}%
\bibitem [{\citenamefont {Kozin}\ and\ \citenamefont
  {Kyriienko}(2019)}]{2019_Kozin_PRL}%
  \BibitemOpen
  \bibfield  {author} {\bibinfo {author} {\bibfnamefont {V.~K.}\ \bibnamefont
  {Kozin}}\ and\ \bibinfo {author} {\bibfnamefont {O.}~\bibnamefont
  {Kyriienko}},\ }\href {\doibase 10.1103/PhysRevLett.123.210602} {\bibfield
  {journal} {\bibinfo  {journal} {Phys. Rev. Lett.}\ }\textbf {\bibinfo
  {volume} {123}},\ \bibinfo {pages} {210602} (\bibinfo {year}
  {2019})}\BibitemShut {NoStop}%
\bibitem [{\citenamefont {Khemani}\ \emph {et~al.}(2016)\citenamefont
  {Khemani}, \citenamefont {Lazarides}, \citenamefont {Moessner},\ and\
  \citenamefont {Sondhi}}]{2016_Khemani_PRL}%
  \BibitemOpen
  \bibfield  {author} {\bibinfo {author} {\bibfnamefont {V.}~\bibnamefont
  {Khemani}}, \bibinfo {author} {\bibfnamefont {A.}~\bibnamefont {Lazarides}},
  \bibinfo {author} {\bibfnamefont {R.}~\bibnamefont {Moessner}}, \ and\
  \bibinfo {author} {\bibfnamefont {S.~L.}\ \bibnamefont {Sondhi}},\ }\href
  {\doibase 10.1103/PhysRevLett.116.250401} {\bibfield  {journal} {\bibinfo
  {journal} {Phys. Rev. Lett.}\ }\textbf {\bibinfo {volume} {116}},\ \bibinfo
  {pages} {250401} (\bibinfo {year} {2016})}\BibitemShut {NoStop}%
\bibitem [{\citenamefont {Huang}\ \emph {et~al.}(2018)\citenamefont {Huang},
  \citenamefont {Wu},\ and\ \citenamefont {Liu}}]{2018_Biao_PRL}%
  \BibitemOpen
  \bibfield  {author} {\bibinfo {author} {\bibfnamefont {B.}~\bibnamefont
  {Huang}}, \bibinfo {author} {\bibfnamefont {Y.-H.}\ \bibnamefont {Wu}}, \
  and\ \bibinfo {author} {\bibfnamefont {W.~V.}\ \bibnamefont {Liu}},\ }\href
  {\doibase 10.1103/PhysRevLett.120.110603} {\bibfield  {journal} {\bibinfo
  {journal} {Phys. Rev. Lett.}\ }\textbf {\bibinfo {volume} {120}},\ \bibinfo
  {pages} {110603} (\bibinfo {year} {2018})}\BibitemShut {NoStop}%
\bibitem [{\citenamefont {Zhang}\ \emph {et~al.}(2017)\citenamefont {Zhang},
  \citenamefont {Hess}, \citenamefont {Kyprianidis}, \citenamefont {Becker},
  \citenamefont {Lee}, \citenamefont {Smith}, \citenamefont {Pagano},
  \citenamefont {Potirniche}, \citenamefont {Potter}, \citenamefont
  {Vishwanath} \emph {et~al.}}]{2017_Vishwanath_Nature}%
  \BibitemOpen
  \bibfield  {author} {\bibinfo {author} {\bibfnamefont {J.}~\bibnamefont
  {Zhang}}, \bibinfo {author} {\bibfnamefont {P.~W.}\ \bibnamefont {Hess}},
  \bibinfo {author} {\bibfnamefont {A.}~\bibnamefont {Kyprianidis}}, \bibinfo
  {author} {\bibfnamefont {P.}~\bibnamefont {Becker}}, \bibinfo {author}
  {\bibfnamefont {A.}~\bibnamefont {Lee}}, \bibinfo {author} {\bibfnamefont
  {J.}~\bibnamefont {Smith}}, \bibinfo {author} {\bibfnamefont
  {G.}~\bibnamefont {Pagano}}, \bibinfo {author} {\bibfnamefont {I.-D.}\
  \bibnamefont {Potirniche}}, \bibinfo {author} {\bibfnamefont {A.~C.}\
  \bibnamefont {Potter}}, \bibinfo {author} {\bibfnamefont {A.}~\bibnamefont
  {Vishwanath}},  \emph {et~al.},\ }\href
  {https://www.nature.com/articles/nature21413} {\bibfield  {journal} {\bibinfo
   {journal} {Nature}\ }\textbf {\bibinfo {volume} {543}},\ \bibinfo {pages}
  {217} (\bibinfo {year} {2017})}\BibitemShut {NoStop}%
\bibitem [{\citenamefont {Autti}\ \emph {et~al.}(2018)\citenamefont {Autti},
  \citenamefont {Eltsov},\ and\ \citenamefont {Volovik}}]{2018_Autti_PRL}%
  \BibitemOpen
  \bibfield  {author} {\bibinfo {author} {\bibfnamefont {S.}~\bibnamefont
  {Autti}}, \bibinfo {author} {\bibfnamefont {V.~B.}\ \bibnamefont {Eltsov}}, \
  and\ \bibinfo {author} {\bibfnamefont {G.~E.}\ \bibnamefont {Volovik}},\
  }\href {\doibase 10.1103/PhysRevLett.120.215301} {\bibfield  {journal}
  {\bibinfo  {journal} {Phys. Rev. Lett.}\ }\textbf {\bibinfo {volume} {120}},\
  \bibinfo {pages} {215301} (\bibinfo {year} {2018})}\BibitemShut {NoStop}%
\bibitem [{\citenamefont {Autti}\ \emph {et~al.}(2021)\citenamefont {Autti},
  \citenamefont {Heikkinen}, \citenamefont {M{\"a}kinen}, \citenamefont
  {Volovik}, \citenamefont {Zavjalov},\ and\ \citenamefont
  {Eltsov}}]{2021_Autti_Nature}%
  \BibitemOpen
  \bibfield  {author} {\bibinfo {author} {\bibfnamefont {S.}~\bibnamefont
  {Autti}}, \bibinfo {author} {\bibfnamefont {P.~J.}\ \bibnamefont
  {Heikkinen}}, \bibinfo {author} {\bibfnamefont {J.~T.}\ \bibnamefont
  {M{\"a}kinen}}, \bibinfo {author} {\bibfnamefont {G.~E.}\ \bibnamefont
  {Volovik}}, \bibinfo {author} {\bibfnamefont {V.~V.}\ \bibnamefont
  {Zavjalov}}, \ and\ \bibinfo {author} {\bibfnamefont {V.~B.}\ \bibnamefont
  {Eltsov}},\ }\href@noop {} {\bibfield  {journal} {\bibinfo  {journal} {Nature
  Materials}\ }\textbf {\bibinfo {volume} {20}},\ \bibinfo {pages} {171}
  (\bibinfo {year} {2021})}\BibitemShut {NoStop}%
\bibitem [{\citenamefont {Smits}\ \emph {et~al.}(2018)\citenamefont {Smits},
  \citenamefont {Liao}, \citenamefont {Stoof},\ and\ \citenamefont {van~der
  Straten}}]{2018_Smits_PRL}%
  \BibitemOpen
  \bibfield  {author} {\bibinfo {author} {\bibfnamefont {J.}~\bibnamefont
  {Smits}}, \bibinfo {author} {\bibfnamefont {L.}~\bibnamefont {Liao}},
  \bibinfo {author} {\bibfnamefont {H.~T.~C.}\ \bibnamefont {Stoof}}, \ and\
  \bibinfo {author} {\bibfnamefont {P.}~\bibnamefont {van~der Straten}},\
  }\href {\doibase 10.1103/PhysRevLett.121.185301} {\bibfield  {journal}
  {\bibinfo  {journal} {Phys. Rev. Lett.}\ }\textbf {\bibinfo {volume} {121}},\
  \bibinfo {pages} {185301} (\bibinfo {year} {2018})}\BibitemShut {NoStop}%
\bibitem [{\citenamefont {Yang}\ and\ \citenamefont
  {Cai}(2021)}]{2021_Cai_PRL}%
  \BibitemOpen
  \bibfield  {author} {\bibinfo {author} {\bibfnamefont {X.}~\bibnamefont
  {Yang}}\ and\ \bibinfo {author} {\bibfnamefont {Z.}~\bibnamefont {Cai}},\
  }\href {\doibase 10.1103/PhysRevLett.126.020602} {\bibfield  {journal}
  {\bibinfo  {journal} {Phys. Rev. Lett.}\ }\textbf {\bibinfo {volume} {126}},\
  \bibinfo {pages} {020602} (\bibinfo {year} {2021})}\BibitemShut {NoStop}%
\bibitem [{\citenamefont {Yue}\ \emph {et~al.}(2022)\citenamefont {Yue},
  \citenamefont {Yang},\ and\ \citenamefont {Cai}}]{2022_Cai_PRB}%
  \BibitemOpen
  \bibfield  {author} {\bibinfo {author} {\bibfnamefont {M.}~\bibnamefont
  {Yue}}, \bibinfo {author} {\bibfnamefont {X.}~\bibnamefont {Yang}}, \ and\
  \bibinfo {author} {\bibfnamefont {Z.}~\bibnamefont {Cai}},\ }\href {\doibase
  10.1103/PhysRevB.105.L100303} {\bibfield  {journal} {\bibinfo  {journal}
  {Phys. Rev. B}\ }\textbf {\bibinfo {volume} {105}},\ \bibinfo {pages}
  {L100303} (\bibinfo {year} {2022})}\BibitemShut {NoStop}%
\bibitem [{\citenamefont {Nie}\ and\ \citenamefont
  {Zheng}(2022)}]{2022_Wei_arXiv}%
  \BibitemOpen
  \bibfield  {author} {\bibinfo {author} {\bibfnamefont {X.}~\bibnamefont
  {Nie}}\ and\ \bibinfo {author} {\bibfnamefont {W.}~\bibnamefont {Zheng}},\
  }\href@noop {} {\bibfield  {journal} {\bibinfo  {journal} {arXiv preprint
  arXiv:2208.10840}\ } (\bibinfo {year} {2022})}\BibitemShut {NoStop}%
\bibitem [{\citenamefont {von Keyserlingk}\ and\ \citenamefont
  {Sondhi}(2016)}]{2016_Keyserlingk_PRB}%
  \BibitemOpen
  \bibfield  {author} {\bibinfo {author} {\bibfnamefont {C.~W.}\ \bibnamefont
  {von Keyserlingk}}\ and\ \bibinfo {author} {\bibfnamefont {S.~L.}\
  \bibnamefont {Sondhi}},\ }\href {\doibase 10.1103/PhysRevB.93.245146}
  {\bibfield  {journal} {\bibinfo  {journal} {Phys. Rev. B}\ }\textbf {\bibinfo
  {volume} {93}},\ \bibinfo {pages} {245146} (\bibinfo {year}
  {2016})}\BibitemShut {NoStop}%
\bibitem [{\citenamefont {Fisher}(1992)}]{1992_Fisher_PRL}%
  \BibitemOpen
  \bibfield  {author} {\bibinfo {author} {\bibfnamefont {D.~S.}\ \bibnamefont
  {Fisher}},\ }\href
  {https://journals.aps.org/prl/abstract/10.1103/PhysRevLett.69.534} {\bibfield
   {journal} {\bibinfo  {journal} {Phys. Rev. Lett.}\ }\textbf {\bibinfo
  {volume} {69}},\ \bibinfo {pages} {534} (\bibinfo {year} {1992})}\BibitemShut
  {NoStop}%
\bibitem [{\citenamefont {Fisher}(1995)}]{1995_Fisher_PRB}%
  \BibitemOpen
  \bibfield  {author} {\bibinfo {author} {\bibfnamefont {D.~S.}\ \bibnamefont
  {Fisher}},\ }\href {\doibase 10.1103/PhysRevB.51.6411} {\bibfield  {journal}
  {\bibinfo  {journal} {Phys. Rev. B}\ }\textbf {\bibinfo {volume} {51}},\
  \bibinfo {pages} {6411} (\bibinfo {year} {1995})}\BibitemShut {NoStop}%
\bibitem [{\citenamefont {Vosk}\ and\ \citenamefont
  {Altman}(2014)}]{2014_Vosk_PRL}%
  \BibitemOpen
  \bibfield  {author} {\bibinfo {author} {\bibfnamefont {R.}~\bibnamefont
  {Vosk}}\ and\ \bibinfo {author} {\bibfnamefont {E.}~\bibnamefont {Altman}},\
  }\href {\doibase 10.1103/PhysRevLett.112.217204} {\bibfield  {journal}
  {\bibinfo  {journal} {Phys. Rev. Lett.}\ }\textbf {\bibinfo {volume} {112}},\
  \bibinfo {pages} {217204} (\bibinfo {year} {2014})}\BibitemShut {NoStop}%
\bibitem [{\citenamefont {Berdanier}\ \emph {et~al.}(2018)\citenamefont
  {Berdanier}, \citenamefont {Kolodrubetz}, \citenamefont {Parameswaran},\ and\
  \citenamefont {Vasseur}}]{2018_Vasseur_PNAS}%
  \BibitemOpen
  \bibfield  {author} {\bibinfo {author} {\bibfnamefont {W.}~\bibnamefont
  {Berdanier}}, \bibinfo {author} {\bibfnamefont {M.}~\bibnamefont
  {Kolodrubetz}}, \bibinfo {author} {\bibfnamefont {S.}~\bibnamefont
  {Parameswaran}}, \ and\ \bibinfo {author} {\bibfnamefont {R.}~\bibnamefont
  {Vasseur}},\ }\href {https://www.pnas.org/doi/full/10.1073/pnas.1805796115}
  {\bibfield  {journal} {\bibinfo  {journal} {Proc. Natl. Acad. Sci.}\ }\textbf
  {\bibinfo {volume} {115}},\ \bibinfo {pages} {9491} (\bibinfo {year}
  {2018})}\BibitemShut {NoStop}%
\bibitem [{\citenamefont {Lazarides}\ \emph {et~al.}(2015)\citenamefont
  {Lazarides}, \citenamefont {Das},\ and\ \citenamefont
  {Moessner}}]{2015_Lazarides_PRL}%
  \BibitemOpen
  \bibfield  {author} {\bibinfo {author} {\bibfnamefont {A.}~\bibnamefont
  {Lazarides}}, \bibinfo {author} {\bibfnamefont {A.}~\bibnamefont {Das}}, \
  and\ \bibinfo {author} {\bibfnamefont {R.}~\bibnamefont {Moessner}},\ }\href
  {\doibase 10.1103/PhysRevLett.115.030402} {\bibfield  {journal} {\bibinfo
  {journal} {Phys. Rev. Lett.}\ }\textbf {\bibinfo {volume} {115}},\ \bibinfo
  {pages} {030402} (\bibinfo {year} {2015})}\BibitemShut {NoStop}%
\bibitem [{\citenamefont {Ponte}\ \emph {et~al.}(2015)\citenamefont {Ponte},
  \citenamefont {Papi\ifmmode~\acute{c}\else \'{c}\fi{}}, \citenamefont
  {Huveneers},\ and\ \citenamefont {Abanin}}]{2015_Ponte_PRL}%
  \BibitemOpen
  \bibfield  {author} {\bibinfo {author} {\bibfnamefont {P.}~\bibnamefont
  {Ponte}}, \bibinfo {author} {\bibfnamefont {Z.}~\bibnamefont
  {Papi\ifmmode~\acute{c}\else \'{c}\fi{}}}, \bibinfo {author} {\bibfnamefont
  {F.~m.~c.}\ \bibnamefont {Huveneers}}, \ and\ \bibinfo {author}
  {\bibfnamefont {D.~A.}\ \bibnamefont {Abanin}},\ }\href {\doibase
  10.1103/PhysRevLett.114.140401} {\bibfield  {journal} {\bibinfo  {journal}
  {Phys. Rev. Lett.}\ }\textbf {\bibinfo {volume} {114}},\ \bibinfo {pages}
  {140401} (\bibinfo {year} {2015})}\BibitemShut {NoStop}%
\bibitem [{\citenamefont {Abanin}\ \emph {et~al.}(2016)\citenamefont {Abanin},
  \citenamefont {{De Roeck}},\ and\ \citenamefont
  {Huveneers}}]{2016_Dmitry_AoP}%
  \BibitemOpen
  \bibfield  {author} {\bibinfo {author} {\bibfnamefont {D.~A.}\ \bibnamefont
  {Abanin}}, \bibinfo {author} {\bibfnamefont {W.}~\bibnamefont {{De Roeck}}},
  \ and\ \bibinfo {author} {\bibfnamefont {F.}~\bibnamefont {Huveneers}},\
  }\href {\doibase https://doi.org/10.1016/j.aop.2016.03.010} {\bibfield
  {journal} {\bibinfo  {journal} {Annals of Physics}\ }\textbf {\bibinfo
  {volume} {372}},\ \bibinfo {pages} {1} (\bibinfo {year} {2016})}\BibitemShut
  {NoStop}%
\bibitem [{\citenamefont {Assaad}\ and\ \citenamefont
  {Herbut}(2013)}]{2013_Assaad_PRX}%
  \BibitemOpen
  \bibfield  {author} {\bibinfo {author} {\bibfnamefont {F.~F.}\ \bibnamefont
  {Assaad}}\ and\ \bibinfo {author} {\bibfnamefont {I.~F.}\ \bibnamefont
  {Herbut}},\ }\href {\doibase 10.1103/PhysRevX.3.031010} {\bibfield  {journal}
  {\bibinfo  {journal} {Phys. Rev. X}\ }\textbf {\bibinfo {volume} {3}},\
  \bibinfo {pages} {031010} (\bibinfo {year} {2013})}\BibitemShut {NoStop}%
\bibitem [{\citenamefont {Girardeau}(1965)}]{1965_Girardeau_Math}%
  \BibitemOpen
  \bibfield  {author} {\bibinfo {author} {\bibfnamefont {M.}~\bibnamefont
  {Girardeau}},\ }\href@noop {} {\bibfield  {journal} {\bibinfo  {journal}
  {Journal of Mathematical Physics}\ }\textbf {\bibinfo {volume} {6}},\
  \bibinfo {pages} {1083} (\bibinfo {year} {1965})}\BibitemShut {NoStop}%
\bibitem [{\citenamefont {Fisher}(1987)}]{1987_Fisher_JAP}%
  \BibitemOpen
  \bibfield  {author} {\bibinfo {author} {\bibfnamefont {D.~S.}\ \bibnamefont
  {Fisher}},\ }\href {https://aip.scitation.org/doi/10.1063/1.338659}
  {\bibfield  {journal} {\bibinfo  {journal} {J. Appl. Phys}\ }\textbf
  {\bibinfo {volume} {61}},\ \bibinfo {pages} {3672} (\bibinfo {year}
  {1987})}\BibitemShut {NoStop}%
\bibitem [{\citenamefont {Efron}(1979)}]{1979_Efron_SIAM}%
  \BibitemOpen
  \bibfield  {author} {\bibinfo {author} {\bibfnamefont {B.}~\bibnamefont
  {Efron}},\ }\href@noop {} {\bibfield  {journal} {\bibinfo  {journal} {SIAM
  review}\ }\textbf {\bibinfo {volume} {21}},\ \bibinfo {pages} {460} (\bibinfo
  {year} {1979})}\BibitemShut {NoStop}%
\bibitem [{\citenamefont {Newman}\ and\ \citenamefont
  {Barkema}(1999)}]{1999_Newman_Oxford}%
  \BibitemOpen
  \bibfield  {author} {\bibinfo {author} {\bibfnamefont {M.~E.}\ \bibnamefont
  {Newman}}\ and\ \bibinfo {author} {\bibfnamefont {G.~T.}\ \bibnamefont
  {Barkema}},\ }\href@noop {} {\emph {\bibinfo {title} {Monte Carlo methods in
  statistical physics}}}\ (\bibinfo  {publisher} {Clarendon Press},\ \bibinfo
  {year} {1999})\BibitemShut {NoStop}%
\end{thebibliography}%


\begin{thebibliography}{3}%
\makeatletter
\providecommand \@ifxundefined [1]{%
 \@ifx{#1\undefined}
}%
\providecommand \@ifnum [1]{%
 \ifnum #1\expandafter \@firstoftwo
 \else \expandafter \@secondoftwo
 \fi
}%
\providecommand \@ifx [1]{%
 \ifx #1\expandafter \@firstoftwo
 \else \expandafter \@secondoftwo
 \fi
}%
\providecommand \natexlab [1]{#1}%
\providecommand \enquote  [1]{``#1''}%
\providecommand \bibnamefont  [1]{#1}%
\providecommand \bibfnamefont [1]{#1}%
\providecommand \citenamefont [1]{#1}%
\providecommand \href@noop [0]{\@secondoftwo}%
\providecommand \href [0]{\begingroup \@sanitize@url \@href}%
\providecommand \@href[1]{\@@startlink{#1}\@@href}%
\providecommand \@@href[1]{\endgroup#1\@@endlink}%
\providecommand \@sanitize@url [0]{\catcode `\\12\catcode `\$12\catcode
  `\&12\catcode `\#12\catcode `\^12\catcode `\_12\catcode `\%12\relax}%
\providecommand \@@startlink[1]{}%
\providecommand \@@endlink[0]{}%
\providecommand \url  [0]{\begingroup\@sanitize@url \@url }%
\providecommand \@url [1]{\endgroup\@href {#1}{\urlprefix }}%
\providecommand \urlprefix  [0]{URL }%
\providecommand \Eprint [0]{\href }%
\providecommand \doibase [0]{http://dx.doi.org/}%
\providecommand \selectlanguage [0]{\@gobble}%
\providecommand \bibinfo  [0]{\@secondoftwo}%
\providecommand \bibfield  [0]{\@secondoftwo}%
\providecommand \translation [1]{[#1]}%
\providecommand \BibitemOpen [0]{}%
\providecommand \bibitemStop [0]{}%
\providecommand \bibitemNoStop [0]{.\EOS\space}%
\providecommand \EOS [0]{\spacefactor3000\relax}%
\providecommand \BibitemShut  [1]{\csname bibitem#1\endcsname}%
\let\auto@bib@innerbib\@empty
\bibitem [{\citenamefont {Terhal}\ and\ \citenamefont
  {DiVincenzo}(2002)}]{2002_Terhal_PRA}%
  \BibitemOpen
  \bibfield  {author} {\bibinfo {author} {\bibfnamefont {B.~M.}\ \bibnamefont
  {Terhal}}\ and\ \bibinfo {author} {\bibfnamefont {D.~P.}\ \bibnamefont
  {DiVincenzo}},\ }\href {\doibase 10.1103/PhysRevA.65.032325} {\bibfield
  {journal} {\bibinfo  {journal} {Phys. Rev. A}\ }\textbf {\bibinfo {volume}
  {65}},\ \bibinfo {pages} {032325} (\bibinfo {year} {2002})}\BibitemShut
  {NoStop}%
\bibitem [{\citenamefont {Xu}\ and\ \citenamefont
  {Deng}(2022)}]{2022_Peng_arXiv}%
  \BibitemOpen
  \bibfield  {author} {\bibinfo {author} {\bibfnamefont {P.}~\bibnamefont
  {Xu}}\ and\ \bibinfo {author} {\bibfnamefont {T.-S.}\ \bibnamefont {Deng}},\
  }\href@noop {} {\bibfield  {journal} {\bibinfo  {journal} {arXiv preprint
  arXiv:2210.15222}\ } (\bibinfo {year} {2022})}\BibitemShut {NoStop}%
\bibitem [{\citenamefont {Fisher}(1992)}]{1992_Fisher_PRL}%
  \BibitemOpen
  \bibfield  {author} {\bibinfo {author} {\bibfnamefont {D.~S.}\ \bibnamefont
  {Fisher}},\ }\href
  {https://journals.aps.org/prl/abstract/10.1103/PhysRevLett.69.534} {\bibfield
   {journal} {\bibinfo  {journal} {Phys. Rev. Lett.}\ }\textbf {\bibinfo
  {volume} {69}},\ \bibinfo {pages} {534} (\bibinfo {year} {1992})}\BibitemShut
  {NoStop}%
\end{thebibliography}%
\bibliographystyle{apsrev4-1}
\end{document}


\renewcommand{\theequation}{S\arabic{equation}}
\renewcommand{\thesection}{S-\arabic{section}}
\renewcommand{\thefigure}{S\arabic{figure}}
\renewcommand{\bibnumfmt}[1]{[S#1]}
\renewcommand{\citenumfont}[1]{S#1}
\setcounter{equation}{0}
\setcounter{figure}{0}

\title{Supplemental Material for: The Sub-Exponential Critical Slowing Down at Floquet Time Crystal Phase Transition}

\author{Wenqian Zhang$^{\blue \dagger}$}
\affiliation{State Key Laboratory of Surface Physics, Key Laboratory of Micro and Nano Photonic Structures (MOE), and Department of Physics, Fudan University, Shanghai 200433, China}

\author{Yadong Wu$^{\blue \dagger}$} 
\affiliation{State Key Laboratory of Surface Physics, Key Laboratory of Micro and Nano Photonic Structures (MOE), and Department of Physics, Fudan University, Shanghai 200433, China}
\affiliation{Shanghai Qi Zhi Institute, AI Tower, Xuhui District, Shanghai 200232, China}

\author{Xingze Qiu}
\affiliation{State Key Laboratory of Surface Physics, Institute of Nanoelectronics and Quantum Computing, and Department of Physics, Fudan University, Shanghai 200438, China}   
\affiliation{Shanghai Qi Zhi Institute, AI Tower, Xuhui District, Shanghai 200232, China}

\author{Jue Nan} 
\email{juenan@fudan.edu.cn}
\affiliation{State Key Laboratory of Surface Physics, Institute of Nanoelectronics and Quantum Computing, and Department of Physics, Fudan University, Shanghai 200438, China} 
\affiliation{Shanghai Qi Zhi Institute, AI Tower, Xuhui District, Shanghai 200232, China}

\author{Xiaopeng Li}
\email{xiaopeng\underline{ }li@fudan.edu.cn}
\affiliation{State Key Laboratory of Surface Physics, Key Laboratory of Micro and Nano Photonic Structures (MOE), and Department of Physics, Fudan University, Shanghai 200433, China}
\affiliation{Institute for Nanoelectronic Devices and Quantum Computing, Fudan University, Shanghai 200433, China}
\affiliation{Shanghai Qi Zhi Institute, AI Tower, Xuhui District, Shanghai 200232, China}
\maketitle

In the first section of this supplemental material, more details on the numerical method of computing dynamics and eigenstates are given. In the second section, we show the relaxation time $\tau$ for the typical value of the coarse grained correlation function $\overline{O(T)}$ defined in Eq. (2) in the main text. We also show the sum of squares due to error (SSE) to demonstrate the quality of fitting $\tau$.

\section{S-1. Free fermion method}
We consider a Floquet time crystal model in a random spin chain in Eq. (1) in the main text. In order to simulate larger systems, we map this model to a system of non-interacting fermions using Jordan-Wigner transformation. We break down the numerical precedure into four parts. For dynamics simulation, we compute the evolution for fermionic operators and map the initial state to its fermionic counterpart. The dynamics of the coarse grained correlation $\overline{O(T)}$ can then be computed in the form of fermionic pfaffians. In terms of resolving eigenstates of the Floquet operator, we define a mapping between quadratic Hamiltonians and coefficient matrices that helps us compute the effective Hamiltnian defined in Eq. (5) in the main text. By standard diagonalization of the effective Hamiltnian, we get the eigenstates of the Floquet operator. In the following, we discuss these step by step.
\subsection{1. Floquet Evolution in the Fermionic Basis}

The Jordan-Wigner transformation reads
\begin{align}
  \hat{\sigma}_j^x &= 2\hat{c}_j^\dag \hat{c}_j -1, \label{JWx}\\
  \hat{\sigma}_j^y &= \prod_{k=1}^{j-1}(1-2\hat{c}_k^\dag \hat{c}_k)\hat{c}_j^\dag +\prod_{k=1}^{j-1}(1-2\hat{c}_k^\dag \hat{c}_k)\hat{c}_j,\label{JWy}\\
  \hat{\sigma}_j^z &= -i\prod_{k=1}^{j-1}(1-2\hat{c}_k^\dag \hat{c}_k)\hat{c}_j^\dag +i\prod_{k=1}^{j-1}(1-2\hat{c}_k^\dag \hat{c}_k)\hat{c}_j,\label{JWz} 
\end{align}
where $\hat{\sigma}_j^\alpha \ (\alpha = x,\ y,\ z)$ is spin-1/2 Pauli operators and $\hat{c}_j\ (\hat{c}_j^\dag)$ represents fermionic annilation (creation) operator. Its inverse transformation is
\begin{align}
    \hat{c}_j &= \frac{1}{2}\prod_{k=1}^{j-1} \left(-\hat{\sigma}_k^x \right) \left( \hat{\sigma}_j^y -i \hat{\sigma}_j^z \right) = \prod_{k=1}^{j-1} \left(-\hat{\sigma}_k^x \right) \hat{\sigma}_j^-, \\
  \hat{c}_j^\dag &= \frac{1}{2} \prod_{k=1}^{j-1} \left(-\hat{\sigma}_k^x \right) \left( \hat{\sigma}_j^y +i \hat{\sigma}_j^z \right) = \prod_{k=1}^{j-1} \left(- \hat{\sigma}_k^x \right) \hat{\sigma}_j^+.
\end{align}

With these transformations, the evolved annilation operator $\hat{c}_m$ $(1<m < L)$ after one Floquet period is
\be
\begin{aligned}
\hat{U}_F^\dag \hat{c}_m \hat{U}_F &= \hat{U}_F^\dag \prod_{k=1}^{m-1}(-\hat{\sigma}_k^x)\hat{\sigma}_m^- \hat{U}_F \\
&=\frac{1}{2}\prod_{k=1}^{m-1}(-\hat{\sigma}_k^x)\hat{\sigma}_m^y\left[\cos(2J_m)e^{-i\pi g\hat{\sigma}_m^x}-i\sin(2J_m)\hat{\sigma}_m^z\hat{\sigma}_{m+1}^z e^{-i \pi g\hat{\sigma}_{m+1}^x}\right]\\
&  \ \ \ \ -\frac{i}{2}\prod_{k=1}^{m-1}(-\hat{\sigma}_k^x)\hat{\sigma}_m^z\left[\cos(2J_{m-1})e^{-i\pi g\hat{\sigma}_m^x}-i\sin(2J_{m-1})\hat{\sigma}_{m-1}^z\hat{\sigma}_m^z e^{-i\pi g\hat{\sigma}_{m-1}^x}\right] \\
&=\frac{i}{2}\sin(2J_m)e^{i\pi g}\hat{c}_{m+1}^\dag -\frac{i}{2}\sin(2J_m)e^{-i\pi g}\hat{c}_{m+1}\\
& \ \ \ \ +\frac{1}{2}\left[\cos(2J_m)-\cos(2J_{m-1})\right]e^{i\pi g}\hat{c}_m^\dag +\frac{1}{2}\left[\cos(2J_{m-1})+\cos(2J_m)\right]e^{-i\pi g}\hat{c}_m\\
& \ \ \ \ -\frac{i}{2}\sin(2J_{m-1})e^{i\pi g}\hat{c}_{m-1}^\dag-\frac{i}{2}\sin(2J_{m-1})e^{-i\pi g}\hat{c}_{m-1},
\end{aligned}
\label{eq:EvolvedAnni}
\ee
where the Floquet operator $\hat{U}_F$ is defined in Eq. (1) in the main text. For convenience we further introduce Majorana fermions through basis transformation 
\begin{equation}
\hat{\gamma}_{2j-1} = -i(\hat{c}_j -\hat{c}_j^\dag) \text{ and }\hat{\gamma}_{2j} = \hat{c}_j + \hat{c}_j^\dag.
\label{DiractoMajorana}
\end{equation}
This transformation can be described by a transformation matrix $U_C$
\begin{equation}
    \left(\hat{\gamma}_1 \hat{\gamma}_2 \cdots \hat{\gamma}_{2L-1} \hat{\gamma}_{2L} \right)^T = U_C \left( \hat{c}_1 \hat{c}_2 \cdots \hat{c}_L \hat{c}_1^\dag \cdots \hat{c}_L^\dag \right)^T.
    \label{Uc}
\end{equation}
In the Majorana basis, Eq.~\eqref{eq:EvolvedAnni} is rewritten as
\begin{align}
\hat{U}_F^\dag \hat{\gamma}_{2m}\hat{U}_F = &\cos(2J_m)\sin(\pi g)\hat{\gamma}_{2m-1}+\cos(2J_m)\cos(\pi g)\hat{\gamma}_{2m} \nonumber\\
&+\sin(2J_m)\cos(\pi g)\hat{\gamma}_{2m+1}-\sin(2J_m)\sin(\pi g)\hat{\gamma}_{2m+2}, \label{EvolvedMajoranaeven}\\
\hat{U}_F^\dag \hat{\gamma}_{2m+1}\hat{U}_F = &-\sin(2J_m)\sin(\pi g)\hat{\gamma}_{2m-1}-\sin(2J_m)\cos(\pi g)\hat{\gamma}_{2m}\nonumber\\
&+\cos(2J_m)\cos(\pi g)\hat{\gamma}_{2m+1}-\cos(2J_m)\sin(\pi g)\hat{\gamma}_{2m+2},\label{EvolvedMajoranaodd}
\end{align}
where $1 \le m \le L-1$. For the boundaries $\left(m = 0 \text{ and } L \right)$, we have
\begin{equation}
    \hat{\gamma}_1 = -\hat{\sigma}_1^z , \
    \hat{\gamma}_{2L} = -i\prod_{k=1}^{L}\left( -\hat{\sigma}_k^x\right) \hat{\sigma}_L^z. \label{gammaedge}
\end{equation}
The evolution of operators $\hat{\gamma}_1$ and $\hat{\gamma}_{2L}$ can be computed with the form Eq.~\eqref{gammaedge}
\be
\hat{U}_F^\dag \hat{\gamma}_1 \hat{U}_F = \cos(\pi g)\hat{\gamma}_1 -\sin(\pi g)\hat{\gamma}_2, \ \hat{U}_F^\dag \hat{\gamma}_{2L}\hat{U}_F = \cos(\pi g)\hat{\gamma}_{2L}+\sin(\pi g)\hat{\gamma}_{2L-1}.
\label{EvolvedMajoranaEdge}
\ee
In sum, after $n$ Floquet periods of evolution, the evolved Majorana fermion $\hat{\gamma}_m (n)\ (1\le m \le 2L)$ can be expressed as
\be
\hat{\gamma}_m(n)=\left(\hat{U}_F^\dag\right)^n \hat{\gamma}_m(0) \left(\hat{U}_F\right)^n = \left[(V_F)^n{\bf{\Gamma}}(0)\right]_m
\label{Evolution}
,
\ee
where the matrix elements of $V_F$ are given in Eq.~\eqref{EvolvedMajoranaeven}, Eq.~\eqref{EvolvedMajoranaodd} and Eq.~\eqref{EvolvedMajoranaEdge}. The vector ${\bf{\Gamma}}$ is defined as 
\begin{equation}
    {\bf{\Gamma}}=(\hat{\gamma}_1, \hat{\gamma}_2, \cdots, \hat{\gamma}_{2L})^T.
    \label{gammabasis}
\end{equation}
From now on, the Floquet dynamics of the model will be described through $V_F$, as shown in subsection 1.3.
\subsection{2. Mapping the Initial N\'eel State to the Respective Fermionic State}
To map the N\'eel state to the corresponding fermionic state, we choose a Hamiltnian $\hat{H}_s$ whose ground state is N\'eel state
\begin{equation}
    \hat{H}_s = \sum_{j=1}^{L-1} J_j \hat{\sigma}_j^z\hat{\sigma}_{j+1}^z.
    \label{anchorHs}
\end{equation}
We would transform $\hat{H}_s$ into the fermionic basis $\hat{H}_f$ and map the fermionic ground state of $\hat{H}_f$ with N\'eel state. Since Hamiltnian $\hat{H}_s$ has two degenerate ground states
\begin{equation}
|+\rangle = \frac{1}{\sqrt{2}}(|\uparrow \downarrow \uparrow \downarrow \cdots \rangle+|\downarrow \uparrow \downarrow \uparrow \cdots \rangle), \ \ 
|-\rangle = \frac{1}{\sqrt{2}}(|\uparrow \downarrow \uparrow \downarrow \cdots \rangle-|\downarrow \uparrow \downarrow \uparrow \cdots \rangle),
\end{equation}
and N\'eel state is the superposition of these two Ising symmetric ground states
\begin{equation}
    |\up \down \up \down \cdots \rangle = \frac{1}{\sqrt{2}}(|+\rangle +|-\rangle),
\end{equation}
we need to map both $|+\rangle$ and $|-\rangle$ to fermionic states before properly mapping N\'eel state. First transform $\hat{H}_s$ to its fermionic counterpart by Eq.~\eqref{JWx}---\eqref{JWz}
\begin{equation}
    \hat{H}_f = \sum_j J_j (-\hat{c}_j^\dag \hat{c}_{j+1}^\dag +\hat{c}_j^\dag \hat{c}_{j+1} -\hat{c}_j \hat{c}_{j+1}^\dag + \hat{c}_j \hat{c}_{j+1}).
    \label{anchorHf}
\end{equation}
We further diagonalize $\hat{H}_f$ through Bogoliubov transformation with the transformation matrix $U_B$
\begin{equation}
    \left(\hat{\eta}_0 \hat{\eta}_1 \cdots \hat{\eta}_{L_1} \hat{\eta}_0^\dag \cdots \hat{\eta}_{L-1}^\dag \right)^T = U_B \left( \hat{c}_1 \hat{c}_2 \cdots \hat{c}_1^\dag \cdots \hat{c}_L^\dag \right)^T.
    \label{Ub}
\end{equation}
This gives the diagonalized Hamiltonian $\hat{H}_{fd} = \sum_k \epsilon_k \hat{\eta}_k^\dag \hat{\eta}_k$, where a constant has been neglected. The degenerate ground states in $\hat{H}_s$ (Eq.~\eqref{anchorHs}) guarantee the existence of zero mode with $\epsilon_0 = 0$ in $\hat{H}_{fd}$. The vacuum state $|0\rangle$ is a ground state and together with $|1\rangle = \hat{\eta}_0^\dag |0\rangle$ can be match with the degenerate ground states $|+\rangle$ and $|-\rangle$ of the spin model
\begin{equation}
|+\rangle = \alpha_+|0\rangle +\beta_+ |1\rangle,\ \  
|-\rangle = \alpha_-|0\rangle +\beta_- |1\rangle. \label{Match}
\end{equation}

The objective of mapping N\'eel state $|\up \down \up \cdots \up \down \rangle$ to a fermionic state can thus be achieved by determining the coefficients in Eq.~\eqref{Match}. Here, $|0\rangle$ is chosen to be a parity-even vacuum. The spin and fermionic representations of the parity operator are shown below. Judging by the fermionic representation, we can always redefine $\hat{\eta}_0$ to satisfy the requirement of the even-parity. Besides parity operator, boundary spin operator is also considered in order to give coefficients $\alpha_+,\ \beta_+,\ \alpha_-,\ \beta_-$.
The first operator we choose is parity operator. We define it in the spin basis and transform to fermions using Eq.~\eqref{JWx}--\eqref{JWz}
\begin{equation}
\hat{P}_x \equiv \hat{\sigma}_1^x \hat{\sigma}_2^x \cdots \hat{\sigma}_L^x = (-i)^{L}\hat{\gamma}_1\hat{\gamma}_2\cdots \hat{\gamma}_{2L}.
\label{parity}
\end{equation}
The spin states $\{ |+\rangle,\ |-\rangle \}$ constitute a 2-dimensional Hilbert subspace. We construct the representation matrix by computing all its elements $\langle \pm| \hat{P}_x |\pm \rangle$. Thus, the representation matrix is
\begin{equation}
    P_x^{(S)} = 
    \begin{pmatrix}
        1& 0 \\
        0& -1
    \end{pmatrix}
\end{equation}
in the spin basis $\left(|+\rangle , \ |-\rangle \right)^T$.
Now we elaborate on the fermionic representation.
We always invoke Wick theorem in computing the expectation value of a string of non-interacting fermionic operators in the vacuum state $|0\rangle$. As an example, the expectation value of the parity operator can be written in the form of a pfaffian
\be
\langle 0| \hat{P}_x |0\rangle = (-i)^{L}\langle 0|\hat{\gamma}_1 \hat{\gamma}_2 \cdots \hat{\gamma}_{2L}|0\rangle ={\rm{pf}}(G),
\label{eq:Parity00}
\ee
where pf({\it G}) denotes the pfaffian of a $2L\times 2L$ antisymmetric matrix $G$. The elements of matrix $G$ are defined as
\be
G_{ij}\equiv\begin{cases} -i\langle 0|\hat{\gamma}_i \hat{\gamma}_j|0\rangle , & i\ne j, \\
  0 ,& i =j.
\end{cases}
\label{eq:GammaCorrelation}
\ee
The expectation value of a quadratic Majorana fermionic operator can be computed through operator transformation. Combining Eq.~\eqref{Uc} and Eq.~\eqref{Ub}, we have the transformation $U_G$ between the Majorana fermionic basis and the quasi-particle basis
\begin{equation}
    \left( \hat{\gamma}_1 \hat{\gamma}_2 \cdots \hat{\gamma}_{2L} \right)^T = U_C U_B^{\dag} \left( \hat{\eta}_0 \hat{\eta}_1 \cdots \hat{\eta}_{L-1} \hat{\eta}_0^\dag \hat{\eta}_1^\dag \cdots \hat{\eta}_{L-1}^\dag \right)^T \equiv U_G \left( \hat{\eta}_0 \hat{\eta}_1 \cdots \hat{\eta}_{L-1} \hat{\eta}_0^\dag \hat{\eta}_1^\dag \cdots \hat{\eta}_{L-1}^\dag \right)^T.
\end{equation}
For $i \ne j$,
\begin{equation}
    \langle 0| \hat{\gamma}_i \hat{\gamma}_j |0\rangle = \sum_{l,k = 1}^L \left(U_G\right)_{il}\langle 0| \hat{\eta}_{l-1} \hat{\eta}_{k-1}^\dag |0\rangle \left( U_G^\dag \right)_{kj} \equiv \left( U_G P U_G^\dag \right)_{ij},
\end{equation}
where the $2L\times 2L$ diagonal matrix $P = \begin{pmatrix} I_L & 0_L \\ 0_L & 0_L\end{pmatrix}$, with $I_L$ denoting an identity matrix.
Note that odd number of fermion operators give vanishing expectation value under vacuum state. This means $\langle1|\hat{P}_x |0\rangle = \langle0|\hat{P}_x|1\rangle = 0$. Finally, for $|1\rangle = \eta_0^\dag |0\rangle$, ground state $|1\rangle$ always has different parity from $|0\rangle$.

By now, we have arrived at the representation matrix of the parity operator
\begin{equation}
    P_x^{(F)} =
    \begin{pmatrix}
        {\rm{pf}}(G) & 0\\
        0& -\rm{pf}(G)
    \end{pmatrix}
\end{equation}
in the fermionic basis $\left ( |0\rangle , \ |1\rangle \right )^T $. Having chosen the parity-even vacuum as $|0\rangle$, the representation matrices have $P_x^{(F)}=P_x^{(S)}$ with ${\rm pf}(G)=1$. The relation of Eq.~\eqref{Match} can be simplified to $\alpha_+=1, \ \alpha_-=\beta_+=0, \ |\beta_-|=1$.

We next turn to operator $\hat{\sigma}^z_1$ to determine $\beta_-$. We denote this operator $\hat{E}_z$ and then map it to Majorana fermions
\begin{equation}
\hat{E}_z \equiv \hat{\sigma}_1^z = -\hat{\gamma}_1.
\end{equation}
The representation of $\hat{E}_z$ is always easy to compute in the spin basis $\left(|+\rangle , \ |-\rangle \right)^T$
\begin{equation}
    E_z^{(S)} = 
    \begin{pmatrix}
        0& 1 \\
        1& 0
    \end{pmatrix}.
    \label{EdgeS}
\end{equation}
The elements of the representation matrix of the $\hat{E}_z$ in the fermionic basis $\left ( |0\rangle , \ |1\rangle \right )^T $ are
\be
\begin{aligned}
  \langle 0| -\hat{\gamma}_1 |0\rangle &= \langle 1|-\hat{\gamma}_1|1\rangle =0, \\
  \langle 1|-\hat{\gamma}_1|0\rangle &=-\langle 0|\hat{\eta}_0 \hat{\gamma}_1 |0\rangle = -C_1^*,\\
  \langle 0|-\hat{\gamma}_1|1\rangle&=-\langle 0|\hat{\gamma}_1 \hat{\eta}^\dag_0 |0\rangle = -C_1,
\end{aligned}
\label{EdgeRep_F}
\ee
where $C_1$ is the first element of vector $\boldsymbol C$ defined as
\begin{equation}
C_j \equiv \langle 0| \hat{\gamma}_j \hat{\eta}_0^\dag |0 \rangle.
\label{vectorC}
\end{equation}
The computation of $\boldsymbol C$ is similar to the matrix $G$, where we invoke the transformation of Majorana fermions
\begin{equation}
    C_j = \langle 0| \hat{\gamma}_j \hat{\eta}_0^\dag |0\rangle = \sum_{k=1}^L\left( U_G \right)_{jk} \langle 0| \hat{\eta}_{k-1}\hat{\eta}_0^\dag |0\rangle = \left( U_G \right)_{j1}.
\end{equation}
Thus $\boldsymbol C$ is the first column of the matrix $U_G$.
Now that
Eq.~\eqref{EdgeRep_F} gives the representation of the edge operator
\begin{equation}
E_z^{(F)} = 
\begin{pmatrix}
    0& -C_1^* \\
    -C_1& 0
\end{pmatrix},
\label{EdgeF}
\end{equation}
we can use the transformation between the two basis (Eq.~\eqref{Match}) to resolve the undetermined coefficient. The two representation Eq.~\eqref{EdgeS} and Eq.~\eqref{EdgeF} have the relation
\begin{equation}
    A^\dag E_z^{(S)} A = E_z^{(F)},
\end{equation}
with the unitary $A=\begin{pmatrix}
    1 & 0 \\
    0 & \beta_-
\end{pmatrix}$,
which gives 
\begin{equation}
    \beta_-=-1/C_1.
\label{Restriction}
\end{equation}
In sum we can now map the N\'eel state to a fermionic state
\be
    |\up \down \up \down \cdots \rangle = \frac{1}{\sqrt{2}}(|+\rangle +|-\rangle) = \frac{1}{\sqrt{2}}(|0\rangle -\frac{1}{C_1}|1\rangle).
    \label{StateMap}
\ee
In principle, once we map a spin state $|+\rangle$ to a fermionic state $|0\rangle$, for any designated state indexed by $k$, we can find an operator $\hat{O}^k$
\begin{equation}
    \hat{O}_s^k |+\rangle \Longleftrightarrow \hat{O}_f^k |0\rangle,
\end{equation}
where spin operator $\hat{O}_s^k$ is related to fermionic operator $\hat{O}_f$ by Jordan Wigner transformation (Eq.~\eqref{JWx}-\eqref{JWz}).
\subsection{3. Dynamics Computation of a Local Spin Operator}
In the main text, we choose the Néel state $|\psi\rangle = |\uparrow \downarrow \uparrow \downarrow \cdots \rangle$ as the initial state. This product state in the z axis simplifies the computation of autocorrelation operator $\hat{\sigma}_j^z(n)\hat{\sigma}_j^z(0)$ into computing expectation value of operator $\hat{\sigma_j^z}$, which composes of the coarse grained order parameter in the main text Eq. (2)
\begin{equation}
\langle \psi| \hat{\sigma}^z_j(n) |\psi\rangle 
=-{\rm{Re}}\left(\langle0|\hat{\sigma}_j^z(n)|1\rangle/C_1\right).
\end{equation}
Here we have used the result of Eq.~\eqref{StateMap}. For $1\le j \le L$, we map the 
\be
\begin{aligned}
\langle 0| \hat{\sigma}_j^z(n)|1\rangle &= -(i)^{j-1} \langle 0| \hat{\gamma}_1(n)\hat{\gamma}_2(n)\hat{\gamma}_3(n) \cdots \hat{\gamma}_{2j-1}(n)\hat{\eta}^\dag_0 (0)|0\rangle \\
&=(i)^{j+1}{\rm{pf}}(M^j(n)),
\end{aligned}
\ee
where we similarly construct a  $2j\times 2j$ antisymmetric matrix $M^j(n)$ ~\cite{2002_Terhal_PRA}
\be
M^j(n) =
\begin{pmatrix}
  iG^j(n) & {\boldsymbol C}^j(n)^T \\
  -{\boldsymbol C}^j(n) & 0
\end{pmatrix}.
\ee
Here the vector ${\boldsymbol C}^j(n)$ and matrix $G^j(n)$ are sections of the larger time dependent ${\boldsymbol C}(n)$ and $G(n)$ whose intial values are defined as Eq.~\eqref{vectorC} and Eq.~\eqref{eq:GammaCorrelation} respectively. Using the evolution of the Majorana fermions in Eq.~\eqref{Evolution}, the vector element $C(n)_j$ is
\begin{equation}
    C(n)_j  = \langle 0| \hat{\gamma}_j (n)\hat{\eta}_0^\dag (0)|0\rangle = \sum_{k=1}^{2L}\left[\left( V_F \right)^n \right]_{jk} \langle 0| \hat{\gamma}_k(0) \hat{\eta}_0^\dag (0) |0\rangle = \sum_{k=1}^{2L}\left[\left( V_F \right)^n \right]_{jk} C_k(0).
\end{equation}
Thus the effect of evolution of vector ${\boldsymbol C}$ is ${\boldsymbol C}(n) = \left(V_F \right)^n {\boldsymbol C}(0)$. In a similar way it can be shown that the time-dependent matrix $G(n) = \left( V_F \right)^n G(0) \left( V_F^\dag \right)^n$. The first $2j$ elements of ${\boldsymbol{C}}(n)$ gives ${\boldsymbol C}^j(n)$ while the first $2j$ rows and first $2j$ columns of $G(n)$ gives $G^j(n)$.

\subsection{4. The Eigenstates of the Floquet Operator}
The eigenstates of the Floquet unitary $\hat{U}_F$ can be achieved by first computing its effective Hamiltonian. Defined by $\exp(-i 2\hat{H}_{\rm eff})\equiv \hat{U}_F$ (Eq. (5) in the main text), it is easier to cope with effective Hamiltonian in the Majorana fermionic basis. From Eq.~\eqref{JWx}---\eqref{JWz} and Eq.~\eqref{DiractoMajorana}, we transform the binary Hamiltnian into 
\be
    \hat{H}(t) = \begin{cases}
    \hat{H}_1\equiv -i \frac{\pi}{2}g\sum_j \gamma_{2j-1}\gamma_{2j},  & 0\le t<t_1, \\
    \hat{H}_2\equiv i\sum_j J_j \gamma_{2j}\gamma_{2j+1} ,& t_1\le t<t_1+t_2.
    \end{cases}
    \label{eq:MajoranaH}
\ee
For operator $\hat{B}$ that is quadratic in $\{\hat{\gamma}_j \}$, define matrix $\tilde{B}$ by
\begin{equation}
\hat{B} = \frac{1}{4}\sum_{ij}(\tilde{B})_{ij}\hat{\gamma}_i \hat{\gamma}_j.
\end{equation}
The mapping from the matrix $\tilde{B}$ to the operator $\hat{B}$ is denoted as $\cal{T}$
\be
    {\cal{T}}(\tilde{B}) = \hat{B}.
\ee
The mapping has the property~\cite{2022_Peng_arXiv}
\be
    [{\cal{T}}(\tilde{H}_1),{\cal{T}}(\tilde{H}_2)] = {\cal{T}}([\tilde{H}_1,\tilde{H}_2]).
\ee
This property gives us the matrix representation of the effective Hamiltonian~\cite{2022_Peng_arXiv}
\be
    \exp(-i\hat{H}_2)\exp(-i\hat{H}_1)=\exp(-i2 \hat{H}_{\rm eff}) \Rightarrow \exp(-i\tilde{H}_2)\exp(-i\tilde{H}_1)=\exp(-i2\tilde{H}_{\rm eff}).
\ee
The effective Hamiltnian is thus $\hat{H}_{\rm eff}=\frac{1}{4}{\bf{\Gamma}}^\dag \tilde{H}_{\rm eff}{\bf{\Gamma}}$, with $\bf{\Gamma}$ defined in Eq.~\eqref{gammabasis}. The eigenstates of the effective Hamiltonian, computed by standard diagonalization, are also the eigenstates of the Floquet operator.

\section{S-2. relaxation Time the in typical Dynamics of coarse grained correlation}
\begin{figure*}
    \begin{center}
        \includegraphics[width=0.7\linewidth]{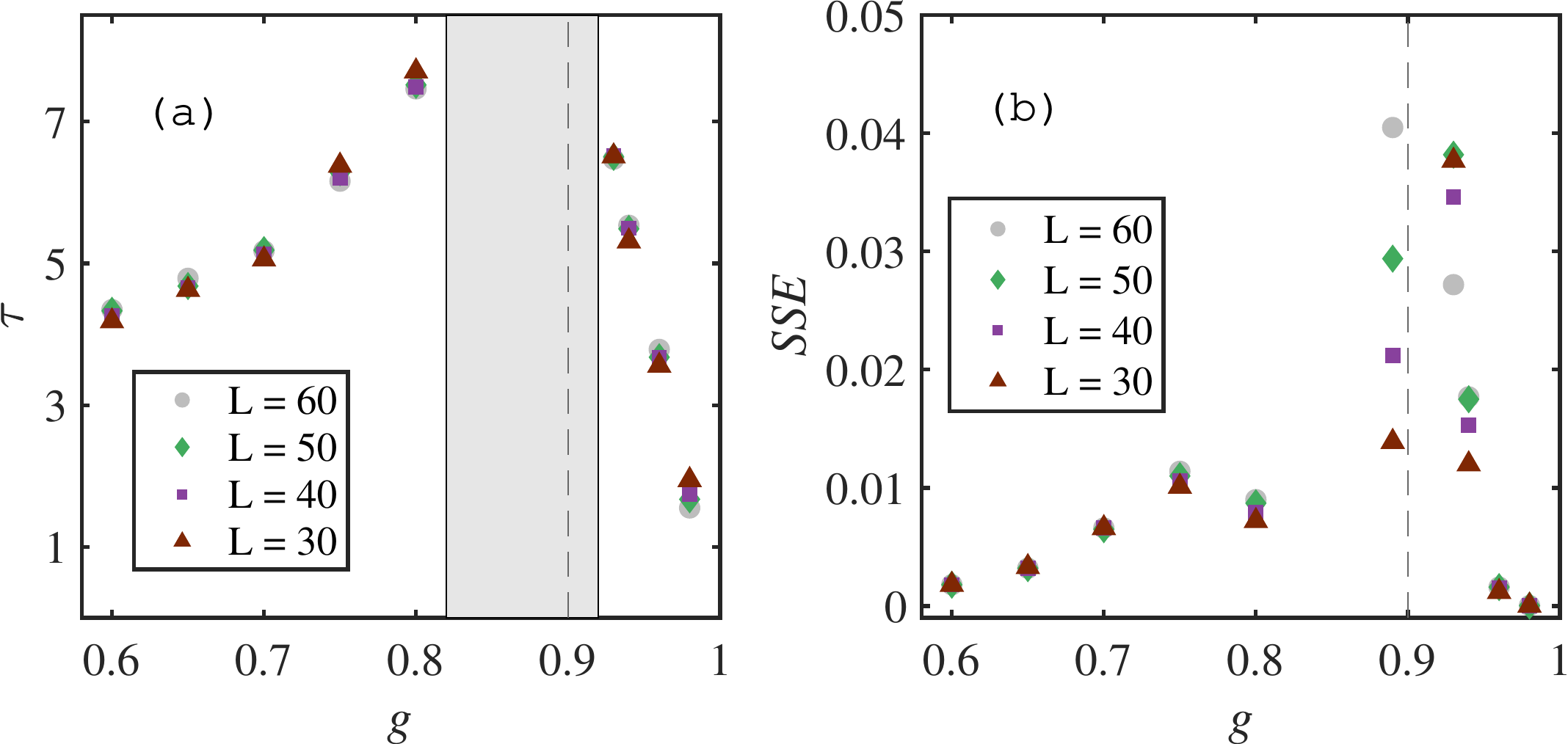}
    \end{center}
    \caption{Extracting relaxation time $\tau$ from the early-time dynamics of the typical order parameter shown in FIG. a (b) of the main text. (a) The relaxation time for different system sizes coincides away from critical point but increases close to the critical point $g_c = 0.9$. The shaded region denotes the area where exponential fitting is not appropriate. (b) The sum of squares due to error ({\it SSE}) of the exponential fit of the typical dynamics shows drastic increase near critical point. The dash lines mark the critical point} 
    \label{fig:1}
\end{figure*}

Previous results have shown the different behaviors from different disorder average~\cite{1992_Fisher_PRL}. 
The relaxation time $\tau$ for $\overline{O(T)}^{\rm{mea}}$ (see Eq. (2)) is shown in the main text. Here we show $\tau$ for the typical dynamics $\overline{O(T)}^{\rm{typ}}$ shown in FIG. 1 (b).  We adopt the fitting function $\overline{O(T)}_{\rm{fit}}=a_1 \exp(-T/\tau)+a_2$. The results are shown in FIG.~\ref{fig:1} (a) up to the system size $L=60$. Similar to the mean case, the relaxation time $\tau$ increases from both sides of the critical point and is insensitive to the system size away from the transition point. The fitting breaks down in the shaded region near the critical point.  This fitting breakdown is shown by the standard fitting quality indicator---sum of squares due to error (${\it SSE}$) (See FIG.~\ref{fig:1} (b)).

$^{\blue \dagger}$ These authors contributed equally to this work. 

\bibliography{supreference}
\bibliographystyle{apsrev4-1}